\documentclass[]{aa}

\usepackage{graphicx}
\usepackage{txfonts}
\usepackage{lipsum}
\usepackage{subcaption}  
\usepackage{lscape}
\usepackage{placeins}
\usepackage{rotating}
\usepackage{tikz}
\usepackage{pdflscape}
\usepackage{xcolor}
\usepackage[colorlinks=true,citecolor=blue,linkcolor=blue,urlcolor=blue]{hyperref}

\begin{document}

\title{The CON-quasar stage of IRAS 07251$-$0248 E}

   \author{Eduardo Gonz\'alez-Alfonso
          \inst{1}
          \and
          Miguel Pereira-Santaella
          \inst{2}
          \and
          Ismael Garc\'{\i}a-Bernete
          \inst{3}
          \and
          Jacqueline Fischer
          \inst{4}
           \and
          Giovanna Speranza
          \inst{2}
         }

   \institute{Universidad de Alcal\'a, Departamento de F\'{\i}sica
     y Matem\'aticas, Campus Universitario, E-28871 Alcal\'a de Henares,
     Madrid, Spain\\
              \email{eduardo.gonzalez@uah.es}
              \and
   Instituto de F\'{\i}sica Fundamental, CSIC, Calle Serrano 123,
                E-28006 Madrid, Spain         
                \and
Centro de Astrobiolog\'{\i}a (CAB), CSIC-INTA, Camino Bajo del
Castillo s/n, Villanueva de la Ca\~nada, E-28692 Madrid, Spain                
                \and
              George Mason University, Department of Physics \& Astronomy,
              MS 3F3, 4400 University Drive, Fairfax, VA 22030, USA
             }

   \authorrunning{Gonz\'alez-Alfonso et al.}
   \titlerunning{The CON-quasar stage of IRAS 07251$-$0248 E}
   

 
   \abstract{ALMA continuum measurements of the local ultra-luminous 
     galaxy IRAS\,07251$-$0248\,E at 667\,$\mu$m
     reveal an extremely compact ($R\lesssim27$\,pc)
     and bright ($T_{\mathrm{B}}\gtrsim200$\,K)
     nucleus with an absorbing foreground envelope
     and a surrounding ($R\sim75$\,pc) disk or torus seen
     nearly face-on. The bright and unresolved nuclear emission implies
     large optical depths ($\tau_{667\mu m}\gtrsim0.5$,
     corresponding to $N_{\mathrm{H}}\gtrsim10^{25}$\,cm$^{-2}$) of hot dust at
     $\gtrsim500$\,K. In addition,
     {\it JWST} observations of the source show
     strong mid-infrared (mid-IR) absorption in the ro-vibrational bands of
     H$_2$O $\nu_2=1-0$ ($5-7$\,$\mu$m) and of other species
     including CO, HCN, C$_2$H$_2$, CH$_4$, and CO$_2$, 
     and {\it Herschel}/PACS observations exhibit strong and
       saturated absorption due to OH, H$_2$O, CH$^+$, and CH.
       We propose a model in which
       the unresolved ALMA submillimeter 
     and {\it JWST} mid-IR continua trace the same nuclear
     source, the former penetrating deep into the nucleus and
     the latter probing the nuclear photosphere.
     The continuum model, which includes trapping of photons
     (the ``greenhouse'' effect),
     indicates that the nuclear ($R_h\approx13$\,pc) luminosity 
     and luminosity surface density are
     $\sim10^{12}\,L_{\odot}$ and
     $\Sigma_{\mathrm{bol}}\approx5\times10^8$\,$L_{\odot}$\,pc$^{-2}$,
     arising from an active galactic nucleus (AGN) so buried that
     high-ionization lines are completely obscured.
     The observed mid-IR gas-phase
     molecular bands probe outflowing gas with velocities of
     $\sim160$\,km\,s$^{-1}$ and are reproduced with 
     the predicted $T_{\mathrm{dust}}$ profile, while the far-IR molecular
     absorption lines are generated in the surrounding
     thick disk or torus with $\tau_{\mathrm{100\,\mu m}}\sim10$.
     We conclude that IRAS\,07251$-$0248 harbors a compact obscured
     nucleus (CON) that hides an AGN currently
       emitting at quasar luminosity. While the
       observed outflow could be driven by radiation pressure,
       we favor the scenario of a (partially) energy-conserving hot bubble 
       caught in a very early phase of
       the expulsion of the
       highly concentrated gas at the galactic nucleus.
}


   \maketitle

\section{Introduction}
\label{intro}

\defcitealias{gon19}{GAS19}

Compact obscured nuclei \citep[CONs, e.g.][]{fal21},
are bright $<100$\,pc galaxy nuclei with extremely high
column densities ($N_{\mathrm{H_2}}\gtrsim10^{25}$\,cm$^{-2}$).
They have been mostly identified via the detection of HCN vibrational
emission in the millimeter \citep[e.g.,][]{aal15,fal19},
by molecular lines 
in absorption in the far-infrared \citep[IR, e.g.,][]{gon04,gon15},
and through the PAH equivalent width method \citep{gber22b,gber25b}.
Their power source is however hard to
establish due to the high nuclear extinction.
Even the luminosity arising from these nuclei is difficult
to establish from millimeter continuum and lines observed with
high-resolution, because trapping of dust emitted photons
increases the dust temperature ($T_d$) within the nucleus
\citep{gon19} \citepalias[hereafter][]{gon19}
and the luminosity depends on $T_d$ at the
nuclear surface (the ``photosphere''), which is better probed
with molecular absorption lines.
Nevertheless, there is growing evidence
that CONs typically harbor an active galactic nucleus
(AGN) based on the extremely compact millimeter
continuum emission as compared with the
more extended gas mass distribution, as observed with ALMA 
\citep{per21}. Still, accurate IR to millimeter continuum models
together with detailed analysis of the hottest
dust mid-IR emission from CONs are scarce.

Here we combine ALMA, {\it James Webb Space Telescope (JWST)},
and {\it Herschel} observations of the eastern (E)
nucleus of IRAS\,07251$-$0248 
to model its continuum emission from the mid-IR
  to submillimeter (submm) wavelengths. IRAS\,07251$-$0248 is a
  merging ultraluminous infrared galaxy (ULIRG) with an
  IR (6-1500\,$\mu$m) luminosity of $10^{12.45}\,L_{\odot}$ and
  nuclear separation of $\sim1.8$\,kpc \citep{lam22};
its E-nucleus dominates the total luminosity, contributing 91\% 
\citep{per21}. There are no reported observations of HCN vibrational
emission from the galaxy, but there are other significant
indications that it harbors a CON from the
strong absorption in the OH\,65\,$\mu$m doublet \citep{gon15},
the 6.2/3.3\,$\mu$m PAH equivalent-width ratio  \citep{gber25b},
and the absorption below the continuum in the blueshifted wing
of the CO 2-1 line \citep{lam22}.
IRAS\,07251$-$0248 is very faint in the {\it Chandra} soft X-ray
band ($0.5-2$\,keV) and undetected in the hard bands of
{\it Chandra} \citep[$2-7$\,keV,][]{iwa11} and {\it NuSTAR}
\citep[$10-24$\,keV,][]{ric21}.
On the other hand, {\it JWST} observations have recently shown
that the source has a nuclear cosmic-ray dominated chemistry
as derived from the high abundances of H$_3^+$ \citep{per24},
hydrocarbons \citep{gber25}, and molecular cations
\citep{spe25}.
We adopt luminosity and angular
distances of $D_L=400$\,Mpc and $D_A=339$\,Mpc
(scale of $1.64\,\mathrm{pc/mas}$),
respectively.

\begin{figure*}[h]
   \centering
\includegraphics[width=17.0cm]{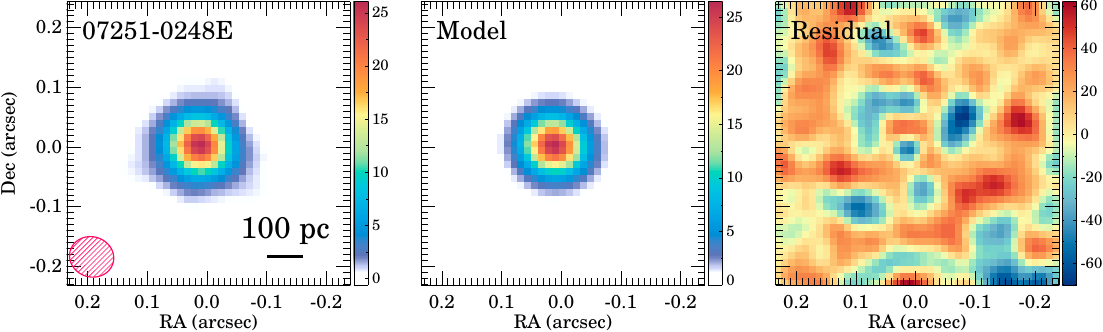}
\caption[]{ALMA 667\,$\mu$m continuum emission of  
  IRAS~07251$-$0248\,E and model. {\it Left}: Observed emission,
  with colored scale in units of mJy/beam. {\it Middle}:
  model, composed of an unresolved source
  ($\mathrm{FWHM}<21$\,mas $=34$\,pc) and a Gaussian
  source ($\mathrm{FWHM}=89\times86\,\mathrm{mas}^2$
  $=146\times141$\,pc$^2$). 
  {\it Right}: residuals in $\mu$Jy/beam.
}
\label{almacont}
\end{figure*} 

\begin{figure*}[h]
   \centering
\includegraphics[width=17.0cm]{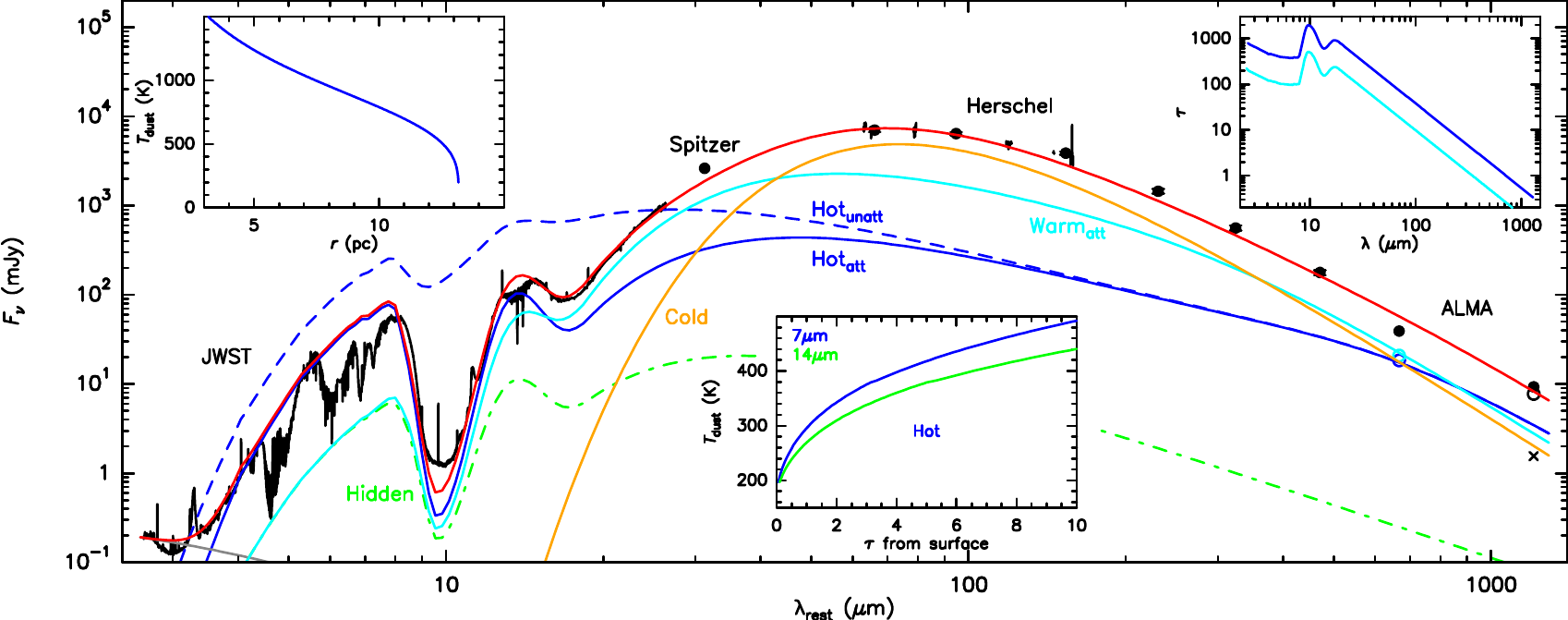}
\caption[]{Spectral energy distribution (SED) of IRAS~07251$-$0248\,E
  from $\lambda_{\mathrm{rest}}=2.6$ to 1200\,$\mu$m and model.
  In black, the full {\it JWST} NIRSpec \& MIRI/MRS
  spectrum is shown, together with
  Spitzer, {\it Herschel}/PACS (from 60 to 160\,$\mu$m, both photometric
  and spectroscopic) and SPIRE (from 230 to 470\,$\mu$m),
  and ALMA (667 and 1200\,$\mu$m) data. The colored circles at 667\,$\mu$m
  indicate the contributions by the unresolved (blue) and resolved
  (light-blue) components inferred from the model in
  Fig.~\ref{almacont}. At 1.2\,mm, the open black circle and the cross indicate
  the flux densities of the unresolved ($<82$\,pc) and resolved 
  ($540\times330$\,pc$^2$) components, respectively \citep{per21}.
  The model includes the $H_C$ in blue (dashed: unattenuated
  emission; solid: attenuated), the $W_C$ in light-blue
  (extincted), a cold component ($C_C$)
  in orange, the stellar component in gray,
  and the ``hidden'' component
    (whose attenuated emission is added to that of the $W_C$,
    see Section~\ref{sech2oband})
    in dotted-dashed green. Red is total.
  The insert panels show details of the model results.
  {\it Upper left}: $T_{\mathrm{dust}}$ profile of the $H_C$.
  {\it Upper right}: the optical depths $\tau$ of the $H_C$ and $W_C$
  as a function of wavelength. {\it Lower middle}:
  the $T_{\mathrm{dust}}$ profile of the $H_C$ as a function
  of the optical depth from surface at 7 and 14\,$\mu$m.  
}
\label{sed}
\end{figure*} 

\section{Observations and results}
\label{obs}

\subsection{ALMA observations of the 667\,$\mu$m continuum and model}
\label{almaobs}

\begin{table}[h]
  \caption{Best-fit parameters for the nuclear 667\,$\mu$m emission of
    IRAS~07251$-$0248\,E using a point-source $+$ Gaussian model}
\label{tab:cont}
\centering
\begin{small}
\begin{tabular}{lcccccccc}
\hline \hline
Parameter & Value & Units \\
\hline
R.A.\,\tablefootmark{a} & 7h27m37.5368s   &  \\
Dec.\,\tablefootmark{a} & $-$2d54m54.267s &  \\
$\chi^2$ & 1.20 \\
\hline
\multicolumn{3}{c}{Point-source}\\
\hline
Flux density & 18.4 $\pm$ 1.0 & mJy \\
\hline
\multicolumn{3}{c}{Gaussian}\\
\hline
FWHM major\,\tablefootmark{b} & 89 $\pm$ 4 & mas \\
FWHM minor\,\tablefootmark{b} & 86 $\pm$ 5 & mas \\
P.A.\tablefoottext{c} & $\cdots$ &  \\
Flux density & 20.8 $\pm$ 0.7 & mJy \\
\hline
\end{tabular}
\end{small}
\tablefoot{\tablefoottext{a}{Common coordinates for the point-source and Gaussian components in the ICRS reference frame.}\tablefoottext{b}{Deconvolved FWHM of the major and minor axis of the Gaussian component.}\tablefoottext{c}{The position angle (P.A.) is not well constrained for this almost circular Gaussian.}
}
\end{table}

ALMA observations at $\nu_{\mathrm{obs}}=413.11$\,GHz
(Band 8, $\lambda_{\mathrm{rest}}=667\,\mu$m)
of IRAS~07251$-$0248\,E (program 2023.1.00942.S, PI: M.
Pereira-Santaella) were carried out on 2023 November 28$^{\mathrm{th}}$
using the 12\,m array with high angular resolution
($75\times67\,\mathrm{mas}^2=123\times110$\,pc$^2$)
  and sensitivity (0.2\,mJy/beam).
Details of the reduction process are given in Appendix~\ref{cont667}.
The map of the $667\,\mu$m continuum
(Fig.~\ref{almacont}) shows a very bright (peak of $\approx25$\,mJy/beam)
and compact emission. To derive its size and flux,
we used the method described in \cite{per21} and attempted to fit the
observed emission with a point-source, a Gaussian, and a point-source
and a Gaussian (pG).
The best fit, shown in Fig.~\ref{almacont}b, was found for the pG
model with parameters listed in Table~\ref{tab:cont}.
Appendix~\ref{cont667} shows that the point-source and the Gaussian fits 
  can be disregarded.

Owing to the high signal-to-noise ratio of the $667\,\mu$m emission
($>120$), the unresolved source of the pG model could be constrained to
have an upper limit for the deconvolved size of
$\mathrm{FWHM}\leq21\,\mathrm{mas}=34$\,pc, assuming
  a Gaussian light distribution.
Conservatively using the equivalence between a Gaussian
source and a disk of uniform brightness
\citep[$\mathrm{radius}\approx0.8\times\mathrm{FWHM}$,][]{sak08},
an upper limit for the nuclear source radius is
$R\lesssim27$\,pc, yielding a brightness of $T_{\mathrm{B}}\gtrsim190$\,K.
The $667\,\mu$m emission is entirely attributed to dust, with the
optical depth given by
\begin{equation}
  \tau_{667\mu m}=\ln\left(1-\frac{10^{-22}\lambda_0^2D_L^2\,F_{667\mu m}}{2\pi k (1+z) T_d R^2}\right)^{-1}
  =\ln\left(1-\frac{7546\,F_{667\mu m}}{T_d R^2}\right)^{-1},
\label{eq:tau667}  
\end{equation}
where $k$ is the Boltzman constant (in cgs units), $\lambda_0=667$\,$\mu$m,
$D_L=400$\,Mpc, $F_{667\mu m}$ is in mJy, $T_d$ is the dust temperature in K,
and the source radius $R$ is in pc. Using $T_d=500$\,K,
eq.~(\ref{eq:tau667}) gives $\tau_{667\mu m}=0.5$ for 
$R=27$\,pc, and has no solution for $R\leq16$\,pc.
Therefore, the ALMA observations reveal an extremely compact source
with a very high column density of hot dust, and will be denoted to as
the ``hot component'' ($H_C$). 

For the Gaussian (resolved) continuum source, a size of
$\mathrm{FWHM}=89\times86\,\mathrm{mas^2}$ is found, corresponding
to a source (most likely a disk or a torus) seen nearly face-on with
a radius of $R\approx75$\,pc.
Equation~(\ref{eq:tau667}) gives $\tau_{667\mu m}\approx0.26$
for an adopted $T_d=123$\,K (Section~\ref{contin}). This component will be
denoted to as the ``warm component'' ($W_C$).

The spectral energy distribution (SED)
of IRAS~07251$-$0248\,E from near-IR to millimeter wavelengths,
including {\it JWST} (Section~\ref{jwstobs}), {\it Spitzer},
{\it Herschel}/PACS and SPIRE, and ALMA data, is shown in Fig.~\ref{sed}.
At 667\,$\mu$m, we plot separately the flux densities for the
$H_C$ (blue) and $W_C$ (light-blue); the black filled
circle is the sum of both and lies below the
extrapolation of the
{\it Herschel}/SPIRE data indicating that a fraction
of the total 667\,$\mu$m emission from the merger
is resolved out by ALMA.
The continuum measurements
at 1.2\,mm are included \citep[Table~3 in][]{per21}, where the black open
circle indicates the unresolved emission 
($R<82$\,pc, which is large enough to include
the emission from the $H_C$ and
$W_C$) and the cross indicates the resolved emission
($\sim500\times300$\,pc$^2$).

\subsection{{\it JWST}  observations of the mid-IR continuum and molecular bands}
\label{jwstobs}

The {\it JWST} observations of IRAS~07251$-$0248 E,
recently presented by \cite{gber25} and \cite{spe25},
were carried out as part of the
{\it JWST} GO Cycle 2 Large Program \#ID:3368 (P.I. L.~Armus and A.~Evans)
including NIRSpec and MIRI/MRS integral field spectroscopy of the nucleus.
Details of the reduction process are given in Appendix~\ref{reduc}.

The {\it JWST} spectrum (Fig.~\ref{sed}) is characterized by deep absorption
of both solid- and gas-phase bands. Besides the silicate 9.7\,$\mu$m and
18\,$\mu$m features, there are deep absorption features
  due to H$_2$O ice at 3 (stretching), 6 (bending), and 13\,$\mu$m (libration),
  and due to hydrogenated amorphous carbon (a-C:H) grains
at 6.85 and 7.25\,$\mu$m. Superposed on this
absorbed continuum, strong gas-phase absorption bands due to CO,
CO$_2$, H$_2$O, HCN,
and C$_2$H$_2$ are detected, as well as
additional weaker bands mostly associated
with hydrocarbons \citep{gber25} and molecular
cations \citep{spe25}. In this absorption-dominated mid-IR spectrum,
no high ionization lines directly probing an AGN are detected
\citep{spo22}. Here we focus on the H$_2$O $\nu_2=1-0$
band at $\sim5.2-7.0$\,$\mu$m displayed in Fig.~\ref{h2oband},
although the HCN, C$_2$H$_2$, CH$_4$, and CO$_2$ bands are also modeled in
Appendix~\ref{hcnc2h2bands}.
The H$_2$O $\nu_2=1-0$ band has been
  previously reported in extragalactic sources by
  \cite{gber24,gber25,gon24,bui24,bui25}.
  We used the H$_2$O spectroscopic parameters 
    tabulated in the HITRAN2020 database \citep{gor22}.

\begin{figure*}[h]
   \centering
\includegraphics[width=18.3cm]{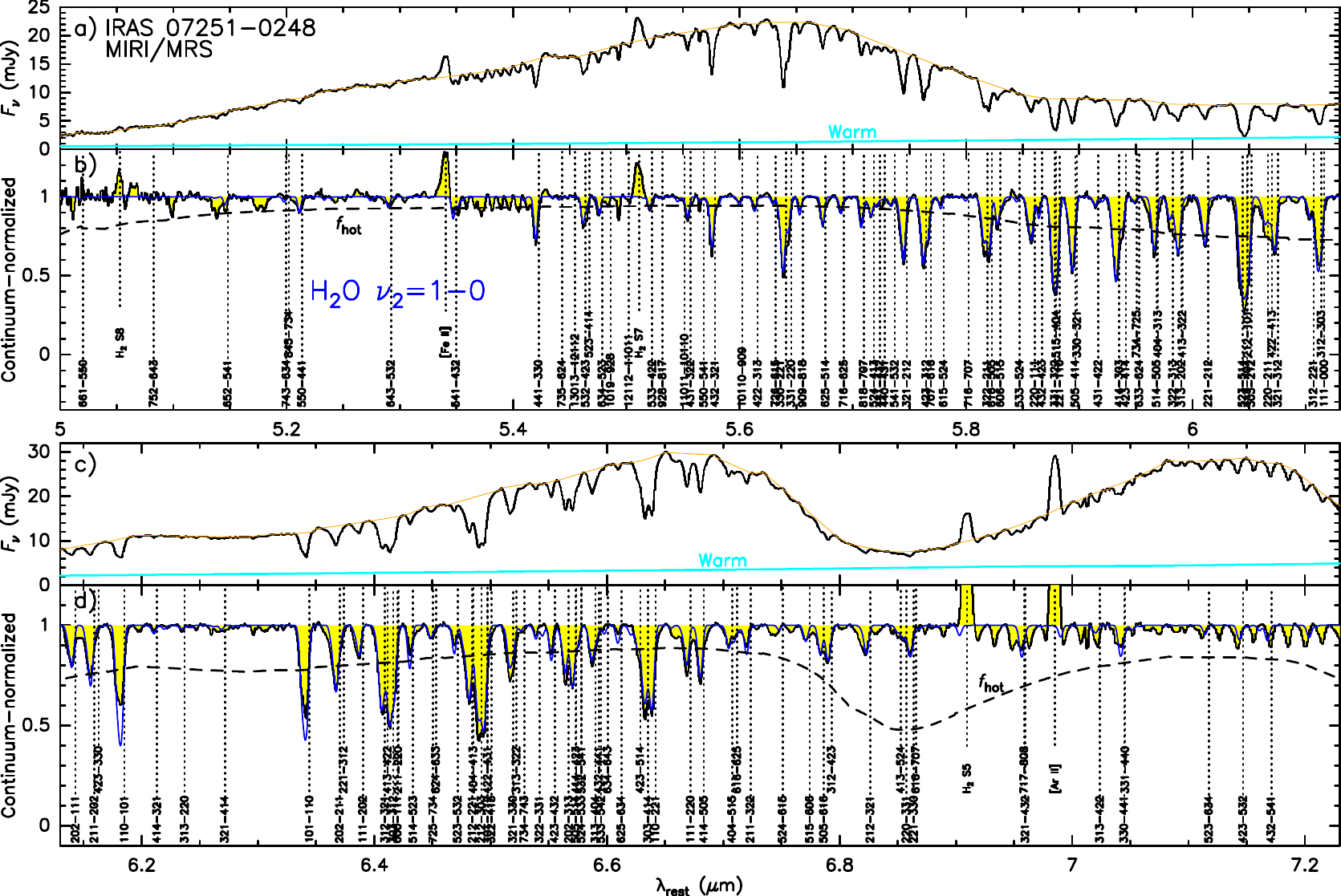}
\caption[]{H$_2$O $\nu_2=1-0$ band in IRAS~07251$-$0248 observed
  with {\it JWST} MIRI/MRS. Panels a and c show the observed
  spectrum, with the orange line indicating the adopted
  continuum level and the light-blue curve showing the
  contribution from the warm component. Panels b and d
  show the continuum normalized spectrum, with the
  blue line showing the model result for the hot component 
  and the dashed line indicating its covering factor
  ($f_{\mathrm{hot}}$). We label the main transitions that potentially 
  contribute to the H$_2$O band as shown in \cite{gon24}, 
  but some very highly-excited lines are not detected in
  IRAS~07251$-$0248. Note that all H$_2$O absorption features are
  blueshifted relative to the labels.
}
\label{h2oband}
\end{figure*} 

The H$_2$O band in IRAS~07251$-$0248 E shows some similarities with, but also
important differences from that observed towards the lower luminosity LIRG
VV\,114\,E\,SW-s2 \citep{gon24}.
In both sources, the bands are in pure absorption
\citep[in contrast with that observed in the LIRG II Zw96-D1, where the P branch
  is in emission;][]{gber24}. However, the continuum-normalized absorption
due to low-excitation H$_2$O lines in IRAS~07251$-$0248 E is stronger
than in VV~114 (see, e.g., the blended feature at $\approx6.05$\,$\mu$m
absorbing $\approx70$\% of the continuum in IRAS~07251$-$0248 but only
$\approx35$\% in VV~114). The opposite happens with the very high-excitation
H$_2$O lines, which are strong in VV~114 but most of them
(above $\sim1500$\,K) undetected in IRAS~07251$-$0248
(see also Fig.~\ref{h2oprof}), indicating significantly
lower excitation in this ULIRG \citep[$\sim200$\,K,][]{gber25}. 
Interestingly, both sources show the H$_2$O lines blueshifted
by $\sim160$\,km\,s$^{-1}$ relative to the emission lines
(Fig.~\ref{redshift}), but the blueshift
in VV~114 increases with higher line excitation while the opposite
happens in IRAS~07251$-$0248 E. Close inspection of the
  H$_2$O line profiles in Figs.~\ref{redshift} and~\ref{h2oprof}
  indicates the presence of two kinematic components, and the
  most blueshifted one (with absorption at velocities up
  to $-500$\,km\,s$^{-1}$) is only seen in the low excitation lines.

\section{A model for the mid-IR to millimeter continuum}
\label{anal}

\subsection{The continuum model}
\label{contin}

Since ALMA observations at 667\,$\mu$m reveal an $H_C$ with
an extreme column density of hot dust
that is expected to emit in the mid-IR,
and {\it JWST} indeed indicates the presence of
  a mid-IR component with a high enough column density
  to generate strong absorption in the molecular bands, 
our hypothesis here is that both observations
trace the same continuum nuclear component
(with $R_h\lesssim27$\,pc as derived from ALMA).
Additional support to this assumption is found in the
CO\,2-1\,230\,GHz rotational line profile displayed
in Fig.~\ref{redshift} \citep{lam22}, which shows
  an asymmetric profile with apparent
  blueshifted absorption between
  $-500$ and $-350$\,km\,s$^{-1}$ falling below the continuum:
  assuming that this absorption is
  produced by the same gas responsible for the low-excitation
  H$_2$O absorption found at similar velocities, the continuum behind
  both CO and H$_2$O line absorption, at $1.3$\,mm and $\sim6$\,$\mu$m,
  should arise from the same physical source. In the following,
we further explore and test this scenario.

The high optical depth of the $H_C$ in the
submm indicates that trapping
of continuum photons will raise $T_d$ within the hot nucleus, and we
have thus generated a grid of ``greenhouse'' models using the method
described in \citetalias{gon19}.
AGN models are used here, together with
an absorption coefficient of dust at $<10$\,$\mu$m similar to that
obtained by \cite{fri11} towards the galactic center and an
absolute calibration in terms of total hydrogen column density
($N_{\mathrm{H}}$, see Appendix~\ref{kabs}) consistent with the
results by \cite{pla11}. In our continuum models, we include the
absorption due to solid-phase silicates, but not due to ices or a:C-H,
so that we aim to fit the envelope of
the $2.6-8$\,$\mu$m continuum (see Fig.~\ref{sed}). For simplicity,
flat density profiles are used.

The grid of greenhouse models for the $H_C$ covers
$\log N_{\mathrm{H}}(\mathrm{cm^{-2}})=24.0-25.6$ and
unattenuated luminosities
$\log L_{\mathrm{unatt}}(L_{\odot})=11.0-12.6$, both with logarithmic
steps of $0.2$\footnote{We use luminosities
integrated from 3 to $1200$\,$\mu$m (basically TIR) because the nuclear component
emits strongly below 10\,$\mu$m, as seen below.}.
A set of mid-IR $2.6-11$\,$\mu$m continuum points unaffected by ice,
a:C-H, or gas-phase bands, together with the flux density at 667\,$\mu$m
of the $H_C$, are used to find the
best-fit model. The unattenuated nuclear emission
  is extinguished by a layer of foreground cold dust, which is
  required because the predicted silicate $9.7$\,$\mu$m
  absorption generated by any relevant greenhouse model is found to be
much shallower than the observed feature.
The fitted parameters are, besides the properties of the
greenhouse model ($N_{\mathrm{H}}$ and $L_{\mathrm{unatt}}$), its radius $R_h$
and the optical depth of the foreground layer evaluated at $6$\,$\mu$m,
$\tau_{6\mu m}^{\mathrm{fg}}$.
Our best-fit model for the $H_C$, with parameters listed 
in Table~\ref{tab}, is shown with a solid-blue line in Fig.~\ref{sed};
the dashed-blue line indicates the unattenuated emission
(i.e., the emission that would be observed without the foreground layer).

\begin{table*}
  \small
  \caption{\label{tab}Properties of the components used to model
    the continuum and the H$_2$O $\nu_2=1-0$ band in IRAS~07251$-$0248 E}
\centering
\begin{tabular}{ccccccccccc}
\hline\hline
Component & $\log L_{\mathrm{unatt}}$\tablefootmark{a} & $T_{d}$\tablefootmark{b}
& $\tau_{\mathrm{6\mu m}}^{\mathrm{int}}$\tablefootmark{c} &
$R$\tablefootmark{d} & $\tau_{6\mu m}^{\mathrm{fg}}$\tablefootmark{e} &
$\log L_{\mathrm{att}}$\tablefootmark{f} & $\log N_{\mathrm{H_2}}$\tablefootmark{g}  &
$X_{\mathrm{H_2O}}$\tablefootmark{h}
& $\log M_{\mathrm{H_2}}$\tablefootmark{i} & $\log L_{\mathrm{int}}$\tablefootmark{j} \\
 & ($L_{\odot}$) & (K) & & (pc) & & ($L_{\odot}$) & (cm$^{-2}$) & ($\times10^{-5}$) & ($M_{\odot}$) & ($L_{\odot}$) \\
\hline
$H_C$ & $12.0(0.10)$ & $\geq200$ & $390(70)$ & $13.2(0.2)$ & $1.16(0.02)$ & $11.4(0.10)$ & $25.3(0.10)$ & $8(2)$ & $8.44(0.10)$ & $11.9-12.2$ \\
$W_C$ & $12.3(0.12)$ & $123(4)$ & $98(22)$ & $71(7)$ & $1.16(0.02)$ & $11.9(0.12)$ & $24.6(0.12)$ & $-$ & $9.13(0.12)$ &  $11.7-12.1$ \\
$C_C$\tablefootmark{k} & $12.1(0.12)$ & $45(5)$ & $-$ & $-$ & 0 & $12.1(0.12)$ & $-$ & $-$ & $9.57(0.12)$ & $11.5-11.8$ \\
\hline
\hline
\end{tabular}
\tablefoot{
\tablefoottext{a}{Unattenuated luminosity ($3-1200$\,$\mu$m) assuming isotropic emission.}
\tablefoottext{b}{Dust temperature. For the $H_C$, radiative
  transfer models compute the $T_d$ profile as shown in the left insert of
  Fig.~\ref{sed}, and the lowest $T_d$ value is here indicated}.
\tablefoottext{c}{Intrinsic optical depth at 6\,$\mu$m of the component;
  the full curves for the $H_C$ and $W_C$ are shown in the right insert
  of Fig.~\ref{sed}.}
\tablefoottext{d}{Radius of the component.}
\tablefoottext{e}{Extinction at 6\,$\mu$m by the foreground layer.}
\tablefoottext{f}{Apparent (attenuated) luminosity ($3-1200$\,$\mu$m) due to foreground
  extinction.}
\tablefoottext{g}{Column density of H$_2$.}
\tablefoottext{h}{H$_2$O abundance relative to H$_2$.}
\tablefoottext{i}{H$_2$ mass derived from the continuum fit (Appendix~\ref{kabs}).}
\tablefoottext{j}{Plausible ranges for the intrinsic luminosities, which only include
    the power sources within the physical regions and consider possible departures from isotropic emission
    (Section~\ref{lumin}).}
\tablefoottext{k}{The cold component is assumed to be optically thin in
    the far-IR, and thus its $\tau_{\mathrm{6\mu m}}^{\mathrm{int}}$ and $R$ cannot be inferred.}
Estimated uncertainties are given in parenthesis.
}
\end{table*}

Once the best fit model for the $H_C$ is obtained,
the $W_C$ is fitted by using {\it JWST} continuum points at
$11.5-26$\,$\mu$m together with the flux density at 667\,$\mu$m of
the resolved component. A graybody (single-$T_d$) extincted
by the same foreground layer as for the $H_C$
  (i.e., with the same $\tau_{6\mu m}^{\mathrm{fg}}$)
is adopted. We add to this graybody an extincted blackbody at 230\,K
primarily contributing at $<10$\,$\mu$m, which has the effect of
producing a better match to the H$_2$O band (see Section~\ref{sech2oband}).
The fitted parameters for the $W_C$ are $T_d$,
  the intrinsic optical depth evaluated at 6\,$\mu$m
  ($\tau_{\mathrm{6\mu m}}^{\mathrm{int}}$), and the source radius $R_w$.
  The resulting ``warm'' continuum, with $T_d=123$\,K, is
  complemented with a ``hidden'' component
  (discussed in Section~\ref{sech2oband}) and 
  shown in light-blue in
Fig.~\ref{sed}. The fitted parameters are listed in
Table~\ref{tab}. Finally, a graybody ``cold'' ($T_d=45$\,K) component
($C_C$) is added to the model to account for the remaining far-IR emission
that cannot be explained with the $H_C$ and $W_C$
(orange curve in Fig.~\ref{sed}).
The $C_C$ also includes the fraction of the
  667\,$\mu$m emission that is resolved out by ALMA.

The fit to the SED of IRAS~07251$-$0248 E indicates that the
mid-IR ($<10$\,$\mu$m) continuum envelope
and the 667\,$\mu$m unresolved emission
can indeed be matched with a single foreground-extincted nuclear component
with unattenuated luminosity $L_{\mathrm{unatt}}\approx10^{12}$\,$L_{\odot}$.
The sizes obtained for this $H_C$ ($R_h=13.2$\,pc) and
  for the $W_C$ ($R_w=71$\,pc)
are consistent with the ALMA 667\,$\mu$m measurements for the
unresolved and resolved components.
Our dust-to-gas mass ratio calibration (Appendix~\ref{kabs}) yields
a gas mass ($M_{\mathrm{H_2}}$ in Table~\ref{tab}) dominated by the $C_C$.
The total $M_{\mathrm{H_2}}=5.3\times10^9$\,$M_{\odot}$ is
comparable with the value obtained from CO\,$1-0$ 
\citep[$4.3\times10^9$\,$M_{\odot}$,][]{gon15}
and CO\,$2-1$ \citep[$5.0\times10^9$\,$M_{\odot}$,][]{lam22}, while the
CO\,$2-1$ value for the central ($r<250$\,pc) region  
is consistently lower \citep[$2.5\times10^9$\,$M_{\odot}$,][]{per21}.
The inserts in Fig.~\ref{sed}
indicate extreme mid-IR optical depths for the $H_C$
  and also for the $W_C$, as well as hot dust
($>500$\,K) within the bulk of the $H_C$.
Nevertheless, the surface
temperature of the $H_C$ is much more moderate
($\approx200-250$\,K) owing to the
nearly complete conversion of the luminosity into blackbody
emission from the nuclear photosphere
($T_d\approx (L_{\mathrm{unatt}}/(4\pi R_h^2 \sigma_{\mathrm{SB}}))^{1/4}=240$\,K).

\subsection{The $6$\,$\mu$m photosphere traced by the H$_2$O $\nu_2=1-0$ band}
\label{sech2oband}

The nuclear $H_C$, with extreme column densities and
  accounting for the bulk of the mid-IR $<8$\,$\mu$m continuum, will naturally
  generate strong mid-IR absorption in the gas-phase molecular bands.
  If the H$_2$O excitation is due to radiative pumping, the H$_2$O $\nu_2=1-0$
  band is key to check whether the inferred $T_d$ profile can
  reproduce the observed line ratios, and will also provide valuable
  information on the gas-phase H$_2$O abundance and on the
  kinematics of the nuclear region.

Using the $T_d$ profile obtained from our best-fit continuum model
for the $H_C$ (Fig.~\ref{sed}),
  we have generated a grid of non-LTE,
non-local radiative transfer models for H$_2$O \citep[e.g.][]{gon24}
by varying only the H$_2$O abundance relative to H$_2$
($X_{\mathrm{H_2O}}$, assumed uniform)
and the gas velocity field. Dust and gas are
uniformly mixed within the
nucleus, and owing to the extreme nuclear mid-IR extinction only
the $6$\,$\mu$m photosphere can be traced with the H$_2$O $\nu_2=1-0$
band.  With the redshift obtained from emission lines
\citep[$z=0.08778$,][see also Fig.~\ref{redshift}]{spe25},
we interpret the blueshift of the H$_2$O lines as an expansion of
the nuclear surface, and apply across the nuclear photosphere an
outflow velocity increasing linearly from 75 to 170\,km\,s$^{-1}$.
Absorption at higher velocities
(up to $\sim-500$\,km\,s$^{-1}$) is also observed in low-excitation lines
(Section~\ref{jwstobs}), and thus an additional H$_2$O shell flowing with
velocities of $150-400$\,km\,s$^{-1}$ was placed in front of the nucleus.

Due to the strong absorption associated with H$_2$O ice around 6\,$\mu$m
and a:C-H grains at 6.85\,$\mu$m, the model predictions are better
compared with the continuum-normalized spectrum in Fig.~\ref{h2oband}b,d.
However, the models that give a good match to most lines 
systematically overestimate the absorption lines located close to
the absorption troughs, particularly at 6.85\,$\mu$m. This specific
model departure indicates the presence of an additional
mid-IR component, unaffected by ice-mantle and H$_2$O
absorption, that dilutes the H$_2$O line
absorption at wavelengths where its strength is comparable to the
observed continuum.
Specifically, the flux density at frequency $\nu$ can be written as
  \begin{equation}
    F_{\nu} = F_{\nu}^{H_C} \, f_{\mathrm{\nu,norm}}^{\mathrm{H_2O}} 
    + \sum_{i\ne H_C} F_{\nu}^{i},  
    \label{fluxdens}
  \end{equation}
where $F_{\nu}^{H_C}$ and $F_{\nu}^{i}$ are the (attenuated) flux densities
from the $H_C$ and all other components $i$  (including the additional mid-IR component),
and $f_{\mathrm{\nu,norm}}^{\mathrm{H_2O}}$
is the continuum-normalized H$_2$O absorption predicted by our model for the $H_C$.
Since the observed continuum $F_{\nu}^{cont}$ has the same expression as eq.~(\ref{fluxdens})
but excluding the $f_{\mathrm{\nu,norm}}^{\mathrm{H_2O}}$ factor, the continuum-normalized
spectrum is given by
\begin{equation}
  \frac{F_{\nu}}{F_{\nu}^{cont}} = 1+ f_{\mathrm{hot}} \left(f_{\mathrm{\nu,norm}}^{\mathrm{H_2O}}-1\right),
  \label{eq:contnorm}
\end{equation}
where we have defined a $\lambda-$dependent covering factor for the $H_C$, which is
the fractional contribution by the $H_C$ to the observed continuum:
\begin{equation}
  f_{\mathrm{hot}} = 1-\frac{\sum_{i\ne H_C} F_{\nu}^{i}}{F_{\nu}^{cont}}
  \label{fhot}
\end{equation}
We use eqs.~(\ref{eq:contnorm}) and (\ref{fhot}) to compare the H$_2$O models
with the observed continuum-normalized spectrum in Fig.~\ref{h2oband}b,d.
At wavelengths where $f_{\mathrm{hot}}$ becomes significantly lower than unity,
because the $H_C$ continuum is severely attenuated by ice or a:C-H and thus
the expected $F_{\nu}^{i}$ from any other component $i$ becomes
comparable to $F_{\nu}^{cont}$, the continuum normalized absorption
due to any H$_2$O line in front of the $H_C$ will be reduced according to
eq.~(\ref{eq:contnorm}). 
The 123\,K component for the $W_C$ in Table~\ref{tab} yields
negligible emission below 10\,$\mu$m, and we find that
a blackbody of 230\,K, attenuated in the same way as the $H_C$ and $W_C$
(green dotted-dashed curve in Fig.~\ref{sed}), 
is required to successfully correct the mentioned discrepancies.
This 230\,K component does not dominate the continuum emission at any
wavelength, so that we will
denote it as the ``hidden'' component. While its origin
is uncertain, we tentatively associate it with the $W_C$ (and add it to the $W_C$;
light blue curve in  Fig.~\ref{sed}). $f_{\mathrm{hot}}$
is shown as black dashed lines in Fig.~\ref{h2oband}b,d.

Our best-fit model for the H$_2$O band (Fig.~\ref{h2oband}b,d) is obtained
for $X_{\mathrm{H_2O}}=8\times10^{-5}$ within the nucleus and
$N_{\mathrm{H_2O}}\approx1.6\times10^{18}$\,cm$^{-2}$ for the
surrounding H$_2$O shell. A more detailed comparison between the
observed line profiles 
and model predictions is shown in Fig.~\ref{h2oprof} and
discussed in Appendix~\ref{h2olines}, but here we emphasize
the following points:
first, that ground-vibrational levels of H$_2$O are excited through the
pumping of the $\nu_2=1$ state by the strong mid-IR radiation that bathes
the nuclear region and its surroundings. Second, that the high
$X_{\mathrm{H_2O}}\sim10^{-4}$ is in line with chemical models of hot
regions where H$_2$O ice has been vaporized and nearly all gas-phase oxygen
not locked into CO is converted into H$_2$O
\citep[e.g.,][]{nom04},
  as well as with models of C-type shocks \citep{kau96b}. And third,
in spite of the very high $T_d$ within the bulk of the $H_C$,
the model captures the observed moderate H$_2$O excitation 
owing to the moderate $T_d$ values of the nuclear photosphere
(lower insert in Fig.~\ref{sed}), which is the
only region the absorption band can probe due to extinction.
We also show in Appendix~\ref{hcnc2h2bands} that the
HCN, C$_2$H$_2$, CH$_4$, and CO$_2$ bands are also reproduced with
our model.

\section{Discussion and conclusions}
\label{disc}

\begin{figure}[h]
   \centering
\includegraphics[width=8cm]{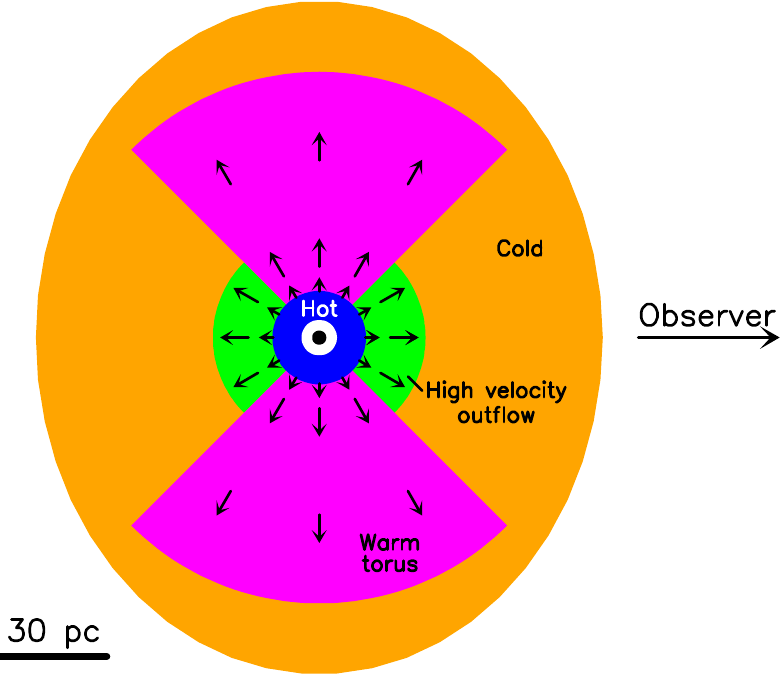}
\caption[]{Schematic representation
  of the model source. The extremely compact and
  expanding core ($H_C$) is shown in blue
  (the 6\,$\mu$m photosphere is its external surface),
  with gas flowing outward  
  (the surrounding shell outflowing at
    $\sim-400$\,km\,s$^{-1}$) in green. The starburst torus shown
  in magenta represents the $W_C$.
  The torus also appears to be expanding, because
    the far-IR lines are blueshifted (Appendix~\ref{farir},
    Fig.~\ref{pacs}).
  Both the $H_C$ and $W_C$ are embedded in an extended
  cold component ($C_C$), in orange.
  The model is axisymmetric about the line
    of sight.
}
\label{sketch}
\end{figure} 

The schematic illustration of the model source
shown in Fig.~\ref{sketch} reflects the
structure inferred from the observations of IRAS~07251$-$0248 E
and model: the expanding nuclear region ($H_C$), is
  surrounded by outflowing gas
(the high velocity foreground shell seen in H$_2$O),
  and a nearly face-on torus ($W_C$).
  The $H_C$ and the $W_C$ are both embedded in an
  envelope representing the $C_C$, which dominates
the far-IR emission between 50 and 300\,$\mu$m. 
In our illustration, the $C_C$ lies in
  front of the $H_C$ and $W_C$, and thus the $C_C$ also represents
the foreground layer that extinguishes
the continuum emission from the $H_C$ and the $W_C$ and
is responsible for the $9.7$ and $18$\,$\mu$m silicate absorption.
  The alternative, that the $W_C$ is a spherical shell
    surrounding the $H_C$, as Fig.~\ref{almacont} might suggest, and is
    thus responsible for the foreground extinction of the $H_C$, is
    slightly disfavored
    because the high optical depth of the $W_C$ (right insert in Fig.~\ref{sed})
    would totally extinguish
    the mid-IR continuum emission from the $H_C$. Nevertheless, it is
    still possible that the high-velocity outflow has cleared the path
    to the $H_C$ along the line of sight and the extinction is caused by
    the residual material of the $W_C$ or by the outflowing material.
    The $C_C$ 
    should include the far-IR emission from the western galaxy of the merger,
    with $9$\% contribution to the overall luminosity
    \citep[$\sim2.5\times10^{11}\,L_{\odot}$,][]{per21}.
Regardless of the origin of the foreground extinction layer, it
will probably have spatially varying chemical
properties as it is fed by outflowing material from
the direction of
the nucleus, generating the observed H$_2$O ice and a:C-H absorption
as the gas-phase H$_2$O freezes out and the hydrocarbons
attach to carbonaceous grains \citep{gber25}.

The $W_C$ is a very optically thick structure
  (right insert in Fig.~\ref{sed}) surrounding the compact nucleus.
  Its far-IR emission is much stronger than the $H_C$ due to its larger
  size, which is similar to the size of the warm and optically thick regions
  that in other ULIRGs generate the far-IR absorption in rotational lines of
  OH, H$_2$O, and other molecular species
  \citep[e.g., Arp~220 and Mrk~231;][]{gon12,gon14b}.  
  The far-IR continuum emission from these regions 
  is usually diluted within the emission from
  colder regions that also contribute to the far-IR,
  which is the reason why (together with re-emission in 
  the line) the far-IR optically thick lines absorb only
  a fraction ($\lesssim20\%$) of the total continuum \citep{gon15}. 
  A similar situation applies to IRAS~07251$-$0248,
  as the $C_C$ dominates the far-IR
  and the continuum from the $W_C$ should be diluted (Fig~\ref{sed}).
  The {\it Herschel}/PACS spectra around the wavelengths of the OH
  doublets at 119, 79, and 65\,$\mu$m, shown in Fig.~\ref{pacs},
  exhibit optically thick absorption in the OH doublets
  and also in lines of H$_2$O, CH$^+$, and CH, so that their
  absorption strengths are 
  sensitive to the fractional contribution of the $W_C$ to the far-IR.
  We show in Appendix~\ref{farir} a
  model for these lines 
  that reasonably fits the absorption troughs, indicating that the modeled 
  $65-120$\,$\mu$m continuum level generated by the $W_C$ is 
  consistent with the available far-IR spectroscopic data.
  All far-IR lines are blueshifted by
    $\sim150$\,km\,s$^{-1}$ (Fig.~\ref{pacs}), indicating
  that the $W_C$ is also expanding.

\subsection{Uncertainties in the inferred parameters}
\label{lumin}

The extreme $N_{\mathrm{H_2}}$ and $M_{\mathrm{H_2}}$ inferred
  in the $H_C$ (Table~\ref{tab}) could still be lower limits, because
  the adopted inner radius of the source is $0.15\times R_{\mathrm{out}}$
  enabling high $T_d>1000$\,K for $r<8$\,pc (left insert in
  Fig.~\ref{sed}). If the dust is however concentrated in a thinner
  shell close to the photosphere, as suggested by the outflow observed
  in the mid-IR molecular bands,
  or if the greenhouse effect in our models
  overestimates the actual $T_d$ values
  \citepalias[see discussion in][]{gon19},
  both $N_{\mathrm{H_2}}$ and $M_{\mathrm{H_2}}$
  would increase because of the overall lower $T_d$. 

As shown in Table~\ref{tab}, the apparent luminosities
  ($L_{\mathrm{att}}$) of the three components yield a total output power
  of $2.3\times10^{12}\,L_{\odot}$, but the combined unattenuated
  luminosities ($L_{\mathrm{unatt}}$) obviously give a higher value.
  This indicates that the values of $L_{\mathrm{unatt}}$ are, for both
  the $W_C$ and the $C_C$, in part re-emission from more inner components,
  i.e. they are not only due to the power sources located within their
  physical regions \citep[e.g.,][]{gon04}. The $W_C$ surrounds
  the $H_C$ and part of its emission will be an effect of the $H_C$
  illumination. Likewise, if the $C_C$ surrounds
  both the $W_C$ and the $H_C$, a fraction of its emission will be
  an effect of heating by the innermost components.
  Defining the ``intrinsic luminosity'' $L_{\mathrm{int}}$ of a
    component as the unattenuated luminosity generated by only the 
    power sources within the physical region, and under the assumption that
    isotropic emission is applicable to the optically thick $H_C$ and
    $W_C$ components, the values of $L_{\mathrm{int}}$ would be $10^{12}$,
    $\sim10^{12}$, and $\sim3\times10^{11}\,L_{\odot}$ for the $H_C$, $W_C$, and $C_C$,
  respectively. These estimates also assume that nearly the entire
  luminosity from the $H_C$ is absorbed and re-emitted by the $W_C$, which
  would then intrinsically generate only half of its $L_{\mathrm{unatt}}$ value;
  the $C_C$ would then produce the remaining emission that
  accounts for the total output power, including
  the emission from the western galaxy.

Nevertheless, isotropic emission must not
necessarily apply to the optically thick
$H_C$ and/or $W_C$ components, in which case their
$L_{\mathrm{int}}$ will depart from the isotropic values quoted above
  \citep[see][]{efs14,efs22}.
  If the morphology of the $W_C$ was a flat disk seen nearly face-on, 
  $L_{\mathrm{unatt}}(W_C)=2\pi R_W^2 \sigma_{\mathrm{SB}}T_d^4\sim10^{12}\,L_{\odot}$
  for $R_W=71$\,pc and $T_d=123$\,K (Table~\ref{tab}).
  The actual scenario is most likely
  intermediate between the spherical and flat morphologies (Fig.~\ref{sketch}).
  Likewise, if the $H_C$ possesses a structure elongated
    along the line of sight, its $L_{\mathrm{int}}$ would exceed $10^{12}\,L_{\odot}$,
    and the $W_C$ emission would be predominantly attributable to external heating.
This possibility is supported by the
fact that, rather than representing a separate physical structure, the $W_C$
appears to be connected to the $H_C$, since both exhibit high column densities
and similar kinematics (Appendix~\ref{farir})
although quite different brightnesses.
In Table~\ref{tab} we provide plausible ranges for 
$L_{\mathrm{int}}$, taking these uncertainties into consideration.

\subsection{A CON hidding a quasar, surrounded by a starburst}
\label{origin}

From a comparative perspective, the high brightness of
the $H_C$ favors the dominant contribution to the luminosity by an AGN.
The luminosity surface density,
$\Sigma_{\mathrm{bol}}\approx4.6\times10^8$\,$L_{\odot}$\,pc$^{-2}$, is
$\sim1$\,dex higher than the values found in
other CONs with comparable sizes and gas mass surface densities
($\Sigma_{\mathrm{H_2}}\approx5\times10^5$\,$M_{\odot}$\,pc$^{-2}$),
such as NGC\,4418
\citep[$(3-7)\times10^7$\,$L_{\odot}$\,pc$^{-2}$,][]{sak13,gon12}, IC\,860
($(1-8)\times10^7$\,$L_{\odot}$\,pc$^{-2}$, \citealp{aal19}; \citetalias{gon19})
or Arp\,220\,W
\citepalias[$\sim3\times10^7$\,$L_{\odot}$\,pc$^{-2}$,][]{gon19}\footnote{Here we
define $\Sigma_{\mathrm{bol}}$ as $L_{\mathrm{bol}}/(4\pi R^2)$, where $R$ is the
radius of the equivalent sphere, and modify the published values in case
that $\Sigma_{\mathrm{bol}}$ is defined in a different way.}.
Our model for the $H_C$ yields an unattenuated SED that
  peaks in the mid-IR (Fig.~\ref{sed}), while the continuum models in
  \citetalias{gon19} consider lower values of $\Sigma_{\mathrm{bol}}$ and the
  corresponding SEDs peak in the far-IR.

Nevertheless, the enormous accumulation of gas in the $H_C$ requires
consideration of the role of a potential transitory burst of star
formation. We first note that the H$_2$O absorption line
shapes do not show any distinct feature at central velocities
(Figs.~\ref{redshift} and \ref{h2oprof}), meaning that
the whole mid-IR photosphere of the $H_C$ is
 outflowing and suggesting the entire
expulsion of the gas from the nuclear region.
If responsible for the outflowing gas,
a nuclear starburst could not have a duration much longer than
the outflow dynamical timescale of $t_{\mathrm{dyn}}\sim R_h/v=8\times10^4$\,yr,
and indeed the free-fall time,
$t_{\mathrm{ff}}= (3\pi/32G\rho)^{1/2}=4.1\times10^4$\,yr (calculated
from the averaged $n_{\mathrm{H_2}}=5.8\times10^5$\,cm$^{-3}$),
accomplishes this constraint. With these short timescales, however,
a pure starburst scenario would imply that most stars are actually in the
protostar phase with the luminosity arising from gas accretion, while
expelling the inflowing material that feeds them.

\cite{tho05} argued that for an optically thick
starburst disk radiating at its Eddington
limit for dust, a characteristic
maximum $\Sigma_{\mathrm{bol}}\sim10^7$\,$L_{\odot}$\,pc$^{-2}$ is attained
that is independent of $\Sigma_{\mathrm{gas}}$
($\approx1.36\times \Sigma_{\mathrm{H_2}}$). This limit for
$\Sigma_{\mathrm{bol}}$ falls too short to explain the inferred value
in the $H_C$, and an AGN would then be responsible for the
bulk of the luminosity.
\cite{and11} postulated the possibility of hot starbursts
with extreme values of
$\Sigma_{\mathrm{bol}}\sim5\times10^8$\,$L_{\odot}$\,pc$^{-2}$ for
$\Sigma_{\mathrm{gas}}\sim5\times10^5$\,$M_{\odot}$\,pc$^{-2}$,
but attributed them to the parsec-scale region fueling a bright AGN.

More recently, \cite{gru18} proposed that for sufficiently high
$\Sigma_{\mathrm{H_2}}>\Sigma_{\mathrm{crit}}\sim3\times10^3$\,$M_{\odot}$\,pc$^{-2}$,
feedback fails due to the strong gravitational force and an extraordinary 
burst of star formation converts most of the gas into stars within a few
$t_{\mathrm{ff}}$. The key feature of these models 
is that the momentum deposition per stellar mass is assumed to be
independent of environment; that is, the ``$\tau_{\mathrm{IR}}$ boost''
characteristic of radiation pressure in optically thick components
is neglected owing to presumable leakage of photons along optically thin paths.
We note that the maximum
$\Sigma_{\mathrm{gas}}\sim1.3\times10^4$\,$M_{\odot}$\,pc$^{-2}$ considered
in their simulations is $\sim40\times$
lower than the value of the $H_C$, but here we assume that their results can be
extrapolated to our more extreme conditions. Adopting a zero-age main
sequence (ZAMS) population with $3\times10^3$\,erg\,s$^{-1}$\,g$^{-1}$,
a stellar mass of $M_*=6.3\times10^8\,M_{\odot}$ is required to account for
$10^{12}\,L_{\odot}$. The current star-formation efficiency would then be
$\mathrm{SFE}=M_*/(M_*+M_{\mathrm{gas}})\sim0.63$ and, with a formation
timescale of $\sim2\,t_{\mathrm{ff}}$, the star-formation rate would be
$\mathrm{SFR}\sim7.7\times10^3$\,$M_{\odot}$\,yr$^{-1}$ with a per-freefall
SFE of $\epsilon_{\mathrm{ff}}\sim0.6$. Within the uncertainties, the
values of SFE and $\epsilon_{\mathrm{ff}}$ agree with results of
\cite{gru18} simulations.

Nevertheless, the above pure-starburst scenario faces serious drawbacks in
the case of the $H_C$ in IRAS~07251$-$0248. 
First, the implied values of $\Sigma_*\sim10^6$\,$M_{\odot}$\,pc$^{-2}$
are found in a few very dense nuclear star clusters (NSCs), but
on smaller spatial scales ($\lesssim5$\,pc) and with lower $M_*$
\citep{gru19,neu20}.
Second, the concentration of the observed gas mass in the $H_C$ would be hard
to explain with such a burst of star formation, because accretion of
$\sim10^9$\,$M_{\odot}$ onto a volume with radius $R_h\sim13$\,pc
would also have to be achieved within $\sim1\,t_{\mathrm{ff}}$
\citep[see also][]{gru19},
requiring unrealistic accretion rates.
Third, protostars do not efficiently generate
the cosmic rays (CRs) required to
explain the observed H$_3^+$, CR-dominated chemistry, and carbonaceous
grain erosion in the $H_C$ \citep{per24,spe25,gber25}.
Fourth, the H$_2$O line profiles strongly suggest a collective motion
of the whole ISM that is better explained by a single feedback event
rather than by the effect of many individual protostars. 
Finally, the ``greenhouse effect'' required to explain the high submm
brightness of the $H_C$ translates into the ``$\tau_{\mathrm{IR}}$ boost'', 
which cannot be neglected for the extreme $\Sigma_{\mathrm{gas}}$ in
the $H_C$, and the observed feedback would be hard to explain
without this effect.

Conversely, an AGN burst responsible for the observed luminosity and
feedback would have enabled smoother gas accretion onto the nuclear
region, owing to the different spatial scales
and solid angles of AGN feedback and
nuclear gas accretion and the stochastic nature
of the former.
While the previous formation of a NSC in IRAS~07251$-$0248 E is a
natural outcome of the highly concentrated ISM,
the extreme $\Sigma_{\mathrm{bol}}$ strongly suggests
a limited star-formation efficiency within a 
compact obscured nucleus (CON) that is powered by an AGN 
currently emitting at quasar level, in agreement
with \cite{per21}.

  With the uncertainties related to geometry in optically thick sources, it
  appears that the $W_C$ is forming stars at high rates equivalent to at least 
  $\sim5\times10^{11}\,L_{\odot}$,
  supporting the view of the composite nature of ULIRGs
  \citep[e.g.,][]{vei09} with the specificity that the 
  $W_C$ is a nuclear starburst characterized by
  high column densities.
  The high gas mass accumulation at the center of the galaxy both
  drives a starburst in situ, and thus the formation
  of a NSC, and efficiently feeds the central supermassive black
  hole \citep[e.g.][]{neu20}, exemplifying the black hole and
  star formation co-evolution in its most extreme form.
  The transition from
  star formation to black hole accretion in the $H_C$ may be smooth in these
  conditions and tidal disruption and capture of 
    (proto)stars close to the black hole may be an efficient black hole 
    growth channel \citep{str09}.

\subsection{Outflow energetics: a hot bubble driving the $H_C$ expansion?}
\label{ener}

With the values of $M_{\mathrm{H_2}}$ and $R$ from Table~\ref{tab}
and a velocity of $160$\,km\,s$^{-1}$, the mass
outflow rate associated with the $H_C$ is estimated as
$\dot{M}=\mu M_{\mathrm{H_2}} v/R=2.9\times10^3$\,$M_{\odot}$\,yr$^{-1}$,
where $\mu=1.36$ accounts for species other than hydrogen.
Here we have assumed 
  that the entire nuclear source is expanding, although we can
  only probe the kinematics of its mid-IR photosphere.
The momentum rate is then
$\dot{P}=\dot{M} v=2.7\times10^{36}\,\mathrm{dyn}=21.5\,L_{\mathrm{AGN}}/c$ and
the energy flux is
$\dot{E}=0.5 \dot{M} v^2=2.05\times10^{43}\,\mathrm{erg\,s^{-1}}=5.4\times10^{-3}\,L_{\mathrm{AGN}}$,
where we have adopted $L_{\mathrm{AGN}}=10^{12}\,L_{\mathrm{\odot}}$
(Table~\ref{tab}).

Radiation pressure on dust grains could in
  principle drive the observed outflow, if it can overcome the
  opposite effect of (self-)gravitation on the gas
  \citep[e.g.,][]{ish15}. Taking
  into account the backpressure effects in optically thick
  environments \citepalias{gon19} and
  with a gas mass of
  $\mu\,M_{\mathrm{H_2}}=3\times10^8$\,$M_{\odot}$ within the
  $H_C$ (Table~\ref{tab}), the gravitational force would however 
  balance the force due to radiation pressure at the $H_C$ surface
  as long as the combined black hole and stellar mass attains  
  $\sim2.5\times10^8$\,$M_{\odot}$. 
  This is plausible, and
  thus the effect of radiation pressure in driving the
  outflow is uncertain.

Another possibility consists of
  an inner hot bubble inflating the nucleus, such that 
  the molecular outflow is generated as the bubble swepts
up the concentrated nuclear ISM. 
We have explored this point using the analytic self-similar solutions
developed by \cite{fau12} (their Appendix~A),
assuming that the hot shocked wind conserves energy.
We first constrain the swept out gas mass to
$\mu\,M_{\mathrm{H_2}}=3\times10^8$\,$M_{\odot}$
at 13.2\,pc with a flat density profile,
and assumed that $1/2$ of the mechanical energy injected by the AGN
($v_{\mathrm{wind}}\,L_{\mathrm{AGN}}/2c$, where $v_{\mathrm{wind}}$ is the inner
wind velocity and $L_{\mathrm{AGN}}=10^{12}\,L_{\odot}$) goes into kinetic
motion of the swept-up gas.
Then, for $v_{\mathrm{wind}}=2\times10^4$\,km\,s$^{-1}$,
a shock velocity driven into the ambient gas of $170$\,km\,s$^{-1}$
at 13.2\,pc is obtained -in agreement with the blueshift of the
peak absorption observed in the molecular
lines. However, the flow time of $\sim5\times10^4$\,yr is about
half the proton cooling time calculated using the two-temperature
effects derived from thermal uncoupling of e$^-$ and protons
\citep{fau12}, and thus a {\it partially} energy-conserving phase
is not ruled out.

This hot bubble scenario is an ultracompact 
version of the (partially) energy-conserving phase of outflows
driven by AGN \citep{tom15,fer15}, with much lower
shock velocities that are a consequence of the enormous
gas masses that are being swept-up at small radii.
The bulk of the gas will not escape
from the potential well of the galaxy, and thus a series of CON phases
are expected to occur over longer timescales, while holes in the
structure will be increasingly opened generating
in later times more extended and faster OH and CO outflows.
The interesting feature of the proposed hot bubble scenario
  is that the inner wind shock presumably accelerates cosmic-ray
particles \citep[e.g.,][]{zub12}, a key ingredient of the
strong H$_3^+$ absorption \citep{per24} and unique carbon-rich \citep{gber25}
and molecular cation \citep{spe25} chemistry observed in IRAS 07251$-$0248\,E.
The strength of linking kinematics and chemistry
  may favor this scenario over radiation pressure as the origin of the
  observed outflow.

  \begin{acknowledgements}
    We thank the referee for the constructive comments
      that helped improve the clarity of the manuscript, and
  acknowledge the DD-ERS teams for developing their observing
  program with a zero--exclusive--access period.
  EG-A thanks the Spanish MICINN for support under projects
  PID2022-137779OB-C41 and PID2023-146667NB-I00.
  MPS acknowledges support under grants RYC2021-033094-I, CNS2023-145506 and
  PID2023-146667NB-I00 funded by MCIN/AEI/10.13039/501100011033 and the
  European Union NextGenerationEU/PRTR.
  IGB is supported by the Programa Atracci\'on de Talento Investigador
  ``C\'esar Nombela'' via grant 2023-T1/TEC-29030 funded by the
  Community of Madrid.  
  This work is based on observations made with the NASA/ESA/CSA James Webb Space
  Telescope. The data were obtained from the Mikulski Archive for Space
  Telescopes at the Space Telescope Science Institute, which is operated by
  the Association of Universities for Research in Astronomy, Inc., under
  NASA contract NAS 5-03127 for JWST; and from the European JWST archive (eJWST)
operated by the ESAC Science Data Centre (ESDC) of the European Space
Agency. These observations are associated with program \#3368.
This paper makes use of the following ALMA data: ADS/JAO.ALMA\#2023.1.00942.S.
ALMA is a partnership of ESO (representing its member states), NSF (USA) and
NINS (Japan), together with NRC (Canada) and NSC and ASIAA (Taiwan) and KASI
(Republic of Korea), in cooperation with the Republic of Chile. The Joint
ALMA Observatory is operated by ESO, AUI/NRAO and NAOJ.
The National Radio Astronomy Observatory is a facility of the National
Science Foundation operated under cooperative agreement by Associated
Universities, Inc.
\end{acknowledgements}

\bibliographystyle{aa}

\begin{appendix}

\section{ALMA observations and models for the 667\,$\mu$m continuum}
\label{cont667}

\begin{figure*}[h]
   \centering
\includegraphics[width=17.0cm]{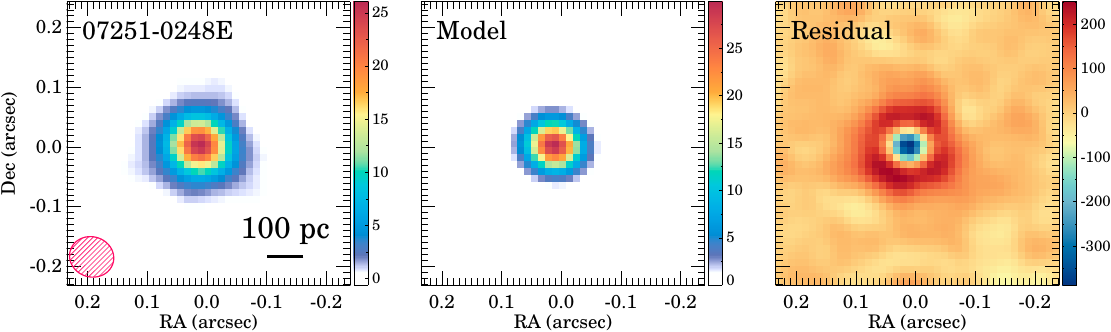}
\includegraphics[width=17.0cm]{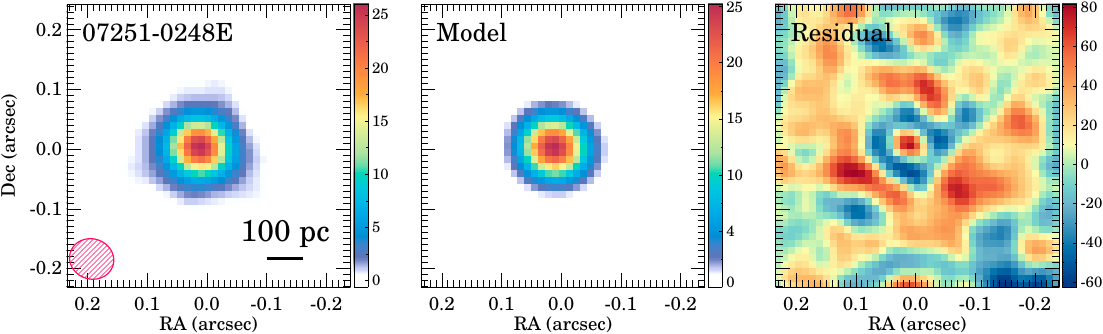}
\caption[]{ALMA 667\,$\mu$m continuum emission of  
  IRAS~07251$-$0248\,E and alternative models: a point source
  (upper) and a Gaussian (lower). 
  {\it Left}: Observed emission,
  with colored scale in units of mJy/beam. {\it Middle}:
  model. {\it Right}: residuals in $\mu$Jy/beam.
  These models give residuals much more prominent than
  our best point-source $+$ Gaussian model fit in
  Fig.~\ref{almacont}.
}
\label{almacontalter}
\end{figure*} 

We obtained high angular resolution ALMA observations of IRAS~07251$-$0248\,E
in Band 8 at $\sim$410\,GHz (rest-frame $\sim$667$\mu$m; program
2023.1.00942.S, PI: M. Pereira-Santaella). Four 1.875\,GHz bandwidth spectral
windows with 10\,km\,s$^{-1}$ channels were defined for these observations.
In this paper, we focus on the continuum detected after combining all the
line-free channels in the spectral windows.
We used the ALMA reduction software CASA (v6.5.4; \citealt{McMullin2007})
to calibrate and clean the data using the standard pipeline. For the
cleaning, we applied the Briggs weighting with a robustness parameter of 1.0.
To improve the quality of the continuum image, we applied a round of phase
self-calibration using the emission from the bright H$_2$O\,448\,GHz line
\citep{per17}. The final continuum image was corrected for the primary
beam. The synthesized beam full-width half-maximum (FWHM) for the combined
continuum image is 75$\times$67\,mas$^2$ and the sensitivity is
0.20\,mJy\,beam$^{-1}$. For Band 8, the absolute flux accuracy is
$\sim$20\% (ALMA Technical Handbook).

Similar to \cite{per21}, we modeled the continuum of IRAS~07251$-$0248\,E
to determine the flux, size, and position of the detected emission.
Briefly, we used three simple models consisting of a point-source,
a Gaussian, and a point-source $+$ Gaussian (pG model).
These models were convolved
with the beam and their parameters varied to minimize the $\chi^2$
obtained from the comparison with the observed image.
Figure\,\ref{almacontalter} shows the results of the fit for
the point-source and Gaussian model, both showing
significant residual structures indicating that they fail to
reproduce the nuclear emission of this object. Instead, the fit is much
better for the pG model, which shows no evident
residuals (Fig.\,\ref{almacont}).

\section{{\it JWST} observations and data reduction}
\label{reduc}

The {\it JWST} observations of IRAS~07251$-$0248\,E have been
described in detail by \cite{gber25} and \cite{spe25}.
Data were taken with the integral-field
spectrograph NIRSpec, using the grating filter pair G395H
\citep[$2.9-5.3$\,$\mu$m with resolution $R\sim2700$;][]{jak22,bok22},
and MIRI MRS
\citep[$4.9-28.1$\,$\mu$m, $R\sim3700-1300$;][]{rie15,wel15,wri15}.
The data were reduced following the standard MRS pipeline
procedure \citep{lab16}, additionally using extra steps to
identify and correct hot and cold pixels \citep{per24,gber24b}.
The spectrum was extracted according with the method described
in \cite{per22} and \cite{gber22,gber24b,gber24}.


\section{The absorption coefficient of dust}
\label{kabs}

\begin{figure}[h]
   \centering
\includegraphics[width=8.5cm]{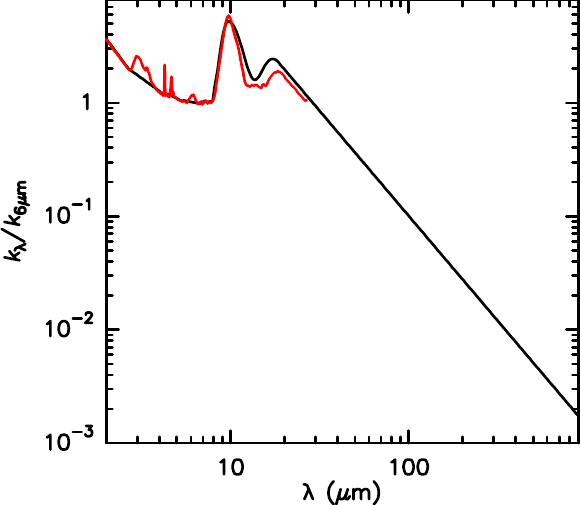}
\caption[]{Adopted absorption coefficient of dust
  normalized at 6\,$\mu$m (black), compared to the
  mid-IR values obtained by \cite{fri11} toward the Galactic
  Center (red).
}
\label{figkabs}
\end{figure} 

Our adopted 6\,$\mu$m-normalized absorption coefficient of dust
from mid-IR to submm wavelengths is shown in Fig.~\ref{figkabs} and compared
with the mid-IR values inferred towards the Galactic Center (GC)
by \cite{fri11}. The latter includes mid-IR features due to ices
and aliphatic hydrocarbons that we have ignored, but otherwise our
curve fits the overall GC curve for $\lambda <10$\,$\mu$m. 
However, we have modified the GC curve for $\lambda=10-20$\,$\mu$m
according to \cite{ros00}, which is in turn based on the work
by \cite{dra84}. This modification gives a better fit to the
IRAS~07251$-$0248 spectrum at the quoted wavelengths. At
$\lambda >20$\,$\mu$m, we use an emissivity index similar to that of the
GC curve ($\beta=1.85$) all the way to millimeter wavelengths.

\cite{pla11} have given an absolute calibration of
$\tau_{250\mu m}/N_{\mathrm{H}}=2.32\times10^{-25}$\,cm$^{2}$, which
for $\beta=1.85$ yields
$\mu \times \mathrm{DGR} \times k_{670\mu m}=2.2\times10^{-2}$\,cm$^{2}$\,g$^{-1}$,
where DGR is the dust-to-gas ratio by mass and $\mu=1.36$ corrects
for all species other than hydrogen.
We have adopted this
calibration to calculate the H$_2$ masses in Table~\ref{tab}, which
roughly agree with independent methods based on CO emission
(Section~\ref{contin}).
This indicates a suitable calibration for at least the $C_C$
component that dominates the gas mass budget, but results are more
uncertain for the $W_C$ and mostly for the $H_C$.
At $6$\,$\mu$m we obtain
$\tau_{6\mu m}/N_{\mathrm{H}}=1.3\times10^{-23}$\,cm$^{2}$,
which is used to calculate the molecular
  abundances in our models where dust and molecules are evenly mixed.
  The hydrogen column density in terms of the optical depth at 667\,$\mu$m is
  $N_{\mathrm{H}}=2.6\times10^{25}\,\tau_{667\mu m}$\,cm$^{-2}$.

\section{Details of the H$_2$O band}
\label{h2olines}

\begin{figure}[h]
   \centering
\includegraphics[width=8.8cm]{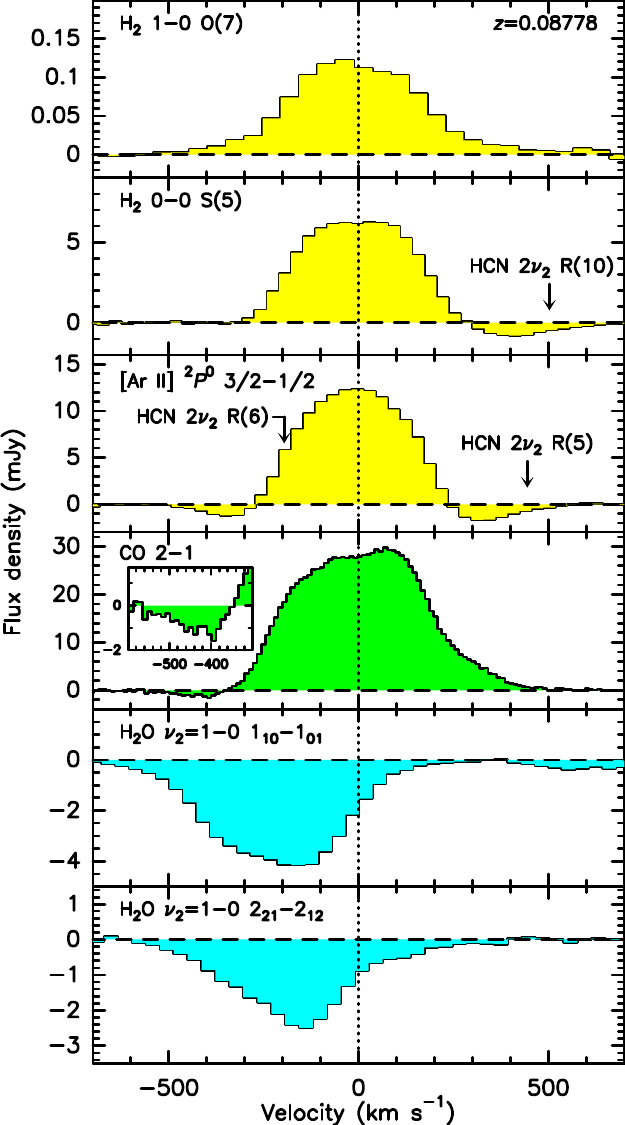}
\caption[]{Line profiles of some emission lines as compared
  with two low-excitation H$_2$O $\nu_2=1-0$ absorption lines.
  The velocity scale on the abscissa is calculated with a
  redshift of $z=0.08778$. All mid-IR molecular 
  absorption lines have peak absorption blueshifted by $100-200$\,km\,s$^{-1}$
  relative to the peak emission lines, as exemplified by the
  H$_2$O lines. Note that the rotational CO 2-1 line shows
  absorption below the continuum at velocities from $-350$ to
  $-570$\,km\,s$^{-1}$ as highlighted in the insert, consistent
  with the H$_2$O wing absorption at similar velocities.
}
\label{redshift}
\end{figure} 

\begin{figure*}[h]
   \centering
\includegraphics[width=18.3cm]{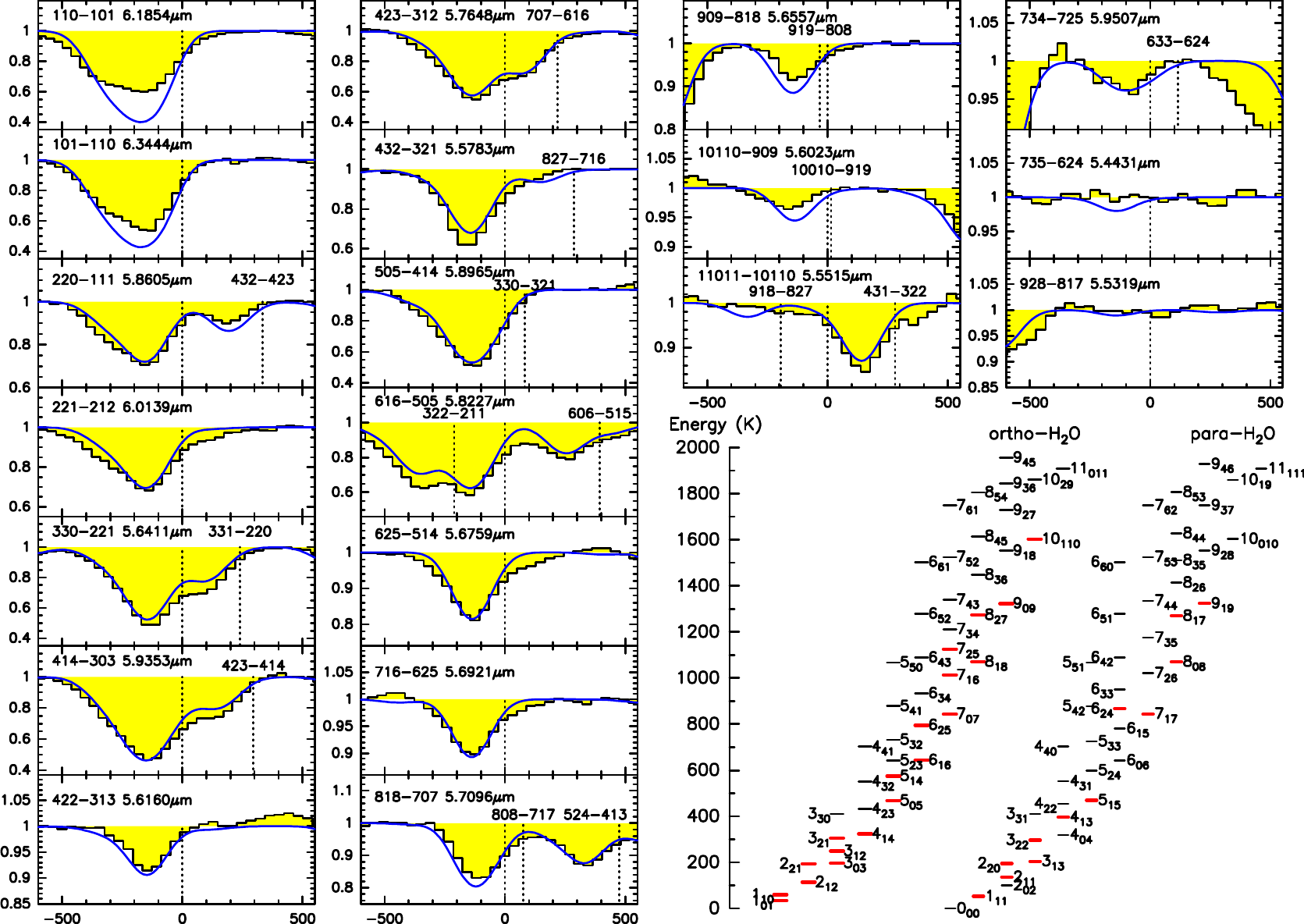}
\caption[]{H$_2$O $\nu_2=1-0$ line profiles in IRAS~07251$-$0248 observed
  with {\it JWST} MIRI/MRS. Filled histograms are the observed profiles
  and blue lines are model results. Abscissa represents velocity
  (in km\,s$^{-1}$) and ordinate is continuum-normalized.
  All detected lines have a peak absorption that is blueshifted by
  $100-200$\,km\,s$^{-1}$.
  The selected transitions are R or Q branch
  lines that are not extremely blended and cover a representative
  range of level energies. Note that some displayed lines are not
  detected (right column).
  The energy level diagram corresponds to the ground state,
  with levels involved in the
  displayed profiles shown in red. 
}
\label{h2oprof}
\end{figure*} 

Figure~\ref{redshift} clearly illustrates the blueshift of the H$_2$O $\nu_2$
absorption lines relative to several emission lines detected with
MIRI/MRS. (Note that the absorption features close to some emission lines
are due to R branch lines of the also blueshifted HCN $2\nu_2$ band.)
The CO 2-1 rotational line 
\citep[see also Fig. C.1 in][]{lam22} shows an asymmetric shape with the
blueshifted part of the profile weaker than the redshifted part. 
In view of the absorption of the continuum detected in the 
velocity range [$-570$,$-350$]\,km\,s$^{-1}$, the asymmetry of the line core
is also attributable to the absorption of the continuum by the outflowing
approaching CO gas, most likely on spatial scales of the $W_C$.
In the lower two panels of Fig.~\ref{redshift}, the profiles correspond
to low-excitation H$_2$O lines that also display absorption up to
$\sim-600$\,km\,s$^{-1}$; other more excited lines do
not show this high-velocity absorption (Fig.~\ref{h2oprof}).

A detailed comparison between a number of observed H$_2$O line shapes
and the best-fit model predictions is presented in Fig.~\ref{h2oprof}.
As shown in the inserted energy level diagram, the displayed lines
(mostly within the R branch) span a broad range of level excitation
in the ground vibrational state, from the ground ortho level ($1_{0,1}$)
to the $10_{1,10}$ one at $\sim1600$\,K, including both backbone and
non-backbone levels.

As described in Section~\ref{sech2oband}, we include two H$_2$O
components along the line of sight to the mid-IR continuum from
the $H_C$: one is evenly mixed with the hot dust responsible
for this continuum, with velocities between 75 and
170\,km\,s$^{-1}$, and the other is a broad layer extending
from the $H_C$ photosphere outwards with velocities between
150 and 400\,km\,s$^{-1}$. In combination with a turbulent velocity
of $90$\,km\,s$^{-1}$, this velocity field can approximately match
the observed profiles.
Radiative pumping fully dominates the excitation of the first component,
so that the H$_2$ density used in the models
($n_{\mathrm{H_2}}=N_{\mathrm{H_2}}/(R_{h}-R_{\mathrm{int}})\sim6\times10^5$\,cm$^{-3}$)
has no impact on the results
as long as the H$_2$O abundance relative to the dust remains unchanged. 
The foreground layer is used to
account for the low-excitation absorption at the highest velocities,
and is shown to produce significant absorption up to the
$5_{0,5}-4_{1,4}$ line ($E_{\mathrm{low}}\approx300$\,K).
In higher excitation lines, the blueshifted absorption is restricted
to velocities $>-300$\,km\,s$^{-1}$ (e.g., $7_{1,6}-6_{2,5}$).
Because of the strong nuclear mid-IR radiation field, the
excitation of this layer component is also dominated by radiative pumping.

In spite of the large column densities of H$_2$O mixed with the dust
in the $H_C$, the predicted absorption troughs are still sensitive to
the H$_2$O abundance relative to the dust, mostly in the highest excitation lines.
This is because the dust opacity restricts the H$_2$O that can
generate absorption to the external layers of the source (i.e. to
$\tau_{6\mu m}\lesssim 1.5$ from the surface), and 
$N(\mathrm{H_2O})\sim5\times10^{18}$\,cm$^{-2}$ in this shell is
much lower than the total value. Our best-fit H$_2$O abundance relative
to H$_2$ of $8\times10^{-5}$ is similar to the value inferred by
\cite{gber25}.

The profiles displayed in Fig.~\ref{h2oprof}
  show that our best fit model overpredicts the absorption in the
low excitation $1_{1,0}-1_{0,1}$ and $1_{0,1}-1_{1,0}$ lines. 
The discrepancy is attributable to potential resonant scattering
of the mid-IR continuum over larger scales, as H$_2$O
is also present in the $W_C$ (Appendix~\ref{farir}).
The model also overpredicts the absorption in some high
excitation lines ($9_{0,9}-8_{1,8}$, $10_{1,10}-9_{0,9}$, $9_{1,8}-8_{2,7}$,
$7_{3,5}-6_{2,4}$), but accounts rather well for the
rest of the lines.

\section{The far-IR molecular absorption}
\label{farir}

\begin{figure}[t]
   \centering
\includegraphics[width=9cm]{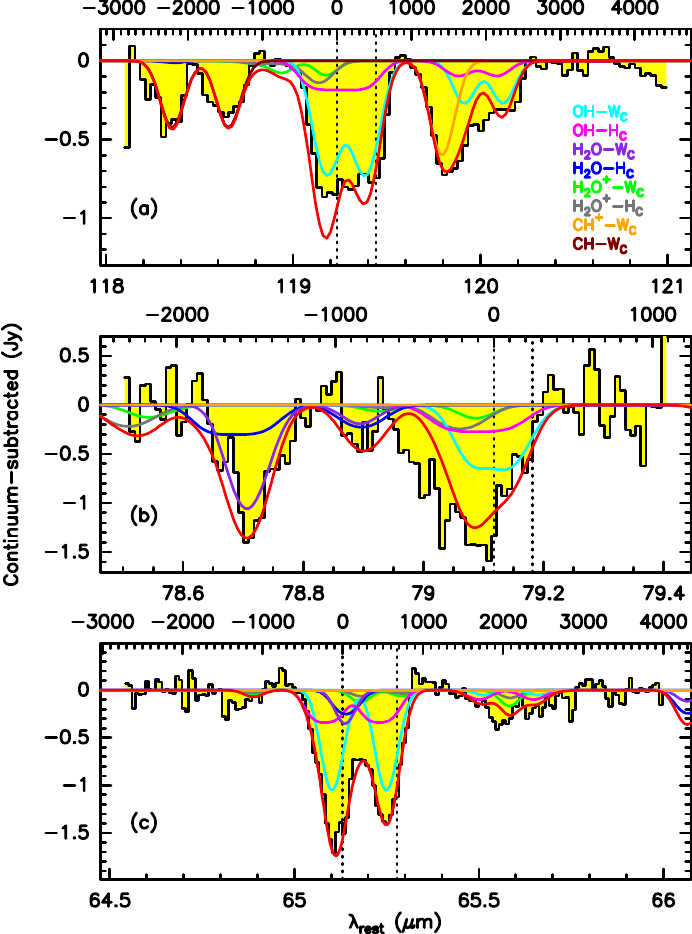}
\caption[]{{\it Herschel}/PACS continuum-subtracted spectra of
  IRAS~07251$-$0248 around 119, 79, and 65\,$\mu$m 
  (filled histograms), and model fit.
  The upper scale in each panel is velocity in km\,s$^{-1}$
    relative to the blue component of each OH doublet, and the two
    vertical dotted lines indicate the positions of the two
    OH $l-$doubling components (rest wavelengths are calculated
    for a redshift $z=0.08778$).
    The contribution to the modeled spectrum by different species
    and components is shown with different colors as labeled in 
    panel (a); red is total.
}
\label{pacs}
\end{figure} 

\begin{table}[t]
  \caption{\label{tab2}Molecular abundances relative to H nuclei
    in the $W_C$ of IRAS~07251$-$0248\,E}
\centering
\begin{tabular}{lc}
  \hline\hline
  Species  &  Abundance\,\tablefootmark{a} \\
\hline
OH         &  $5.0\times10^{-6}$    \\
$^{18}$OH   &  $2.5\times10^{-8}$    \\
H$_2$O     &  $5.0\times10^{-6}$    \\
CH         &  $2.5\times10^{-7}$    \\
CH$^+$     &  $2.5\times10^{-7}$    \\
H$_2$O$^+$ &  $5.0\times10^{-8}$    \\
\hline
\hline
\end{tabular}
\tablefoot{
  \tablefoottext{a}{Abundances
      are inferred from the fit to the far-IR absorption lines
      (Fig.~\ref{pacs}).}
}
\end{table}

Due to the dilution of the $W_C+H_C$ far-IR continuum discussed in
Section~\ref{disc}, we show in Fig.~\ref{pacs} the 
continuum-subtracted far-IR spectra observed with {\it Herschel}/PACS
around the wavelengths of the OH 119, 79, and 65\,$\mu$m doublets.
If the far-IR continuum level predicted by our model
for the $W_C$ is correct, we will be able to match the observed
absorption troughs in this absolute scale.

The far-IR absorption features shown in Fig.~\ref{pacs} are characteristic
of very optically thick regions. In panel a, the ground-state
OH\,119\,$\mu$m doublet exhibits a flat profile with
no dip in the absorption between the two $l-$doubling components.
We also find clear detection of the red component of $^{18}$OH
120.15\,$\mu$m, as well as strong absorption due to CH$^+$\,3-2
at 119.8\,$\mu$m and to the CH\,$(N=3,J=7/2)-(2,5/2)$ doublet around
118.5\,$\mu$m
\citep[$E_{\mathrm{low}}\approx100$\,K; for a level diagram of CH see][]{dav01}.
To our knowledge, this is the only ULIRG where the CH$^+$\,3-2 absorption
is nearly as strong as that due to OH\,119\,$\mu$m.
In panel b, two H$_2$O lines ($4_{2,3}-3_{1,2}$ and $6_{1,5}-5_{2,4}$
at 78.7-78.9\,$\mu$m) are detected, together with a broad feature
that is in part due to the ground-state OH\,79\,$\mu$m doublet but
requires additional absorption at velocities more blueshifted
than $-350$\,km\,s$^{-1}$ from OH.
In panel c, strong absorption in the OH\,65\,$\mu$m doublet
($E_{\mathrm{low}}\approx300$\,K) is seen, which has the second highest
equivalent width among all galaxies observed with {\it Herschel}/PACS
\citep{gon15}. $^{18}$OH at 65.6\,$\mu$m is also detected.
As indicated in Fig.~\ref{pacs} for the OH lines, all features
are blueshifted by $\sim150$\,km\,s$^{-1}$.

Our model for the far-IR lines includes the quoted species
(OH, $^{18}$OH, H$_2$O, CH, CH$^+$)
together with H$_2$O$^+$, which has several rotational lines
in these spectral regions.
Spectroscopic data used for OH, CH, CH$^+$,
  and H$_2$O$^+$ were obtained from  the JPL \citep{pic98} and CDMS
  \citep{mul01,mul05} catalogs.
Since the $W_C$ is very optically thick
($\tau_{100\mu m}\approx10$, Fig.~\ref{sed}), warm
($T_d=123$\,K, Table~\ref{tab}), and has a size ($R_w=71$\,pc) larger
than the $H_C$, it dominates the absorption in the observed lines.
The inferred abundances are listed in Table~\ref{tab2}.
An OH abundance of $\approx5\times10^{-6}$, similar to the value found
in the nuclear region of NGC~4418 \citep[with similar $T_d$,][]{gon12}
is required to generate absorption troughs in the OH\,119 and
65\,$\mu$m doublets comparable to the observed features (light-blue
curves in Fig.~\ref{pacs}). However, the observed very broad OH\,79\,$\mu$m
feature is still underpredicted with just the $W_C$, and we
have thus included in the model the expected OH contribution from the $H_C$,
as well as a model for H$_2$O$^+$ in both components. The 79.1\,$\mu$m
absorption is then better reproduced, although its unpredicted blueshifted
shoulder at 79.0\,$\mu$m suggests that an additional unidentified species
further contributes to the absorption. With a similar abundance of
$\approx5\times10^{-6}$ in the $W_C$, the H$_2$O lines are matched as well,
thus indicating a decline of its abundance of $\sim1$\,dex from the
compact $H_C$ to the more extended $W_C$. We also note the high abundances
inferred for both CH and CH$^+$, suggesting that the hydrocarbon-rich chemistry
observed in the mid-IR towards the nuclear region \citep{gber25} applies
to some extent to the $W_C$.

In summary, the reasonable match to the far-IR absorption lines indicates
that the continuum level of the $W_C$ in the far-IR predicted by our model
is accurate within $\sim20$\%. The similar kinematics found for the far-IR
and mid-IR absorption lines and the high abundances found for CH and CH$^+$
indicate that the physical and chemical processes in the $W_C$ are
also affected by the AGN at the core of the $H_C$.

\section{The HCN, C$_2$H$_2$, CH$_4$, and CO$_2$ bands}
\label{hcnc2h2bands}

\begin{figure*}[h]
   \centering
\includegraphics[width=18.3cm]{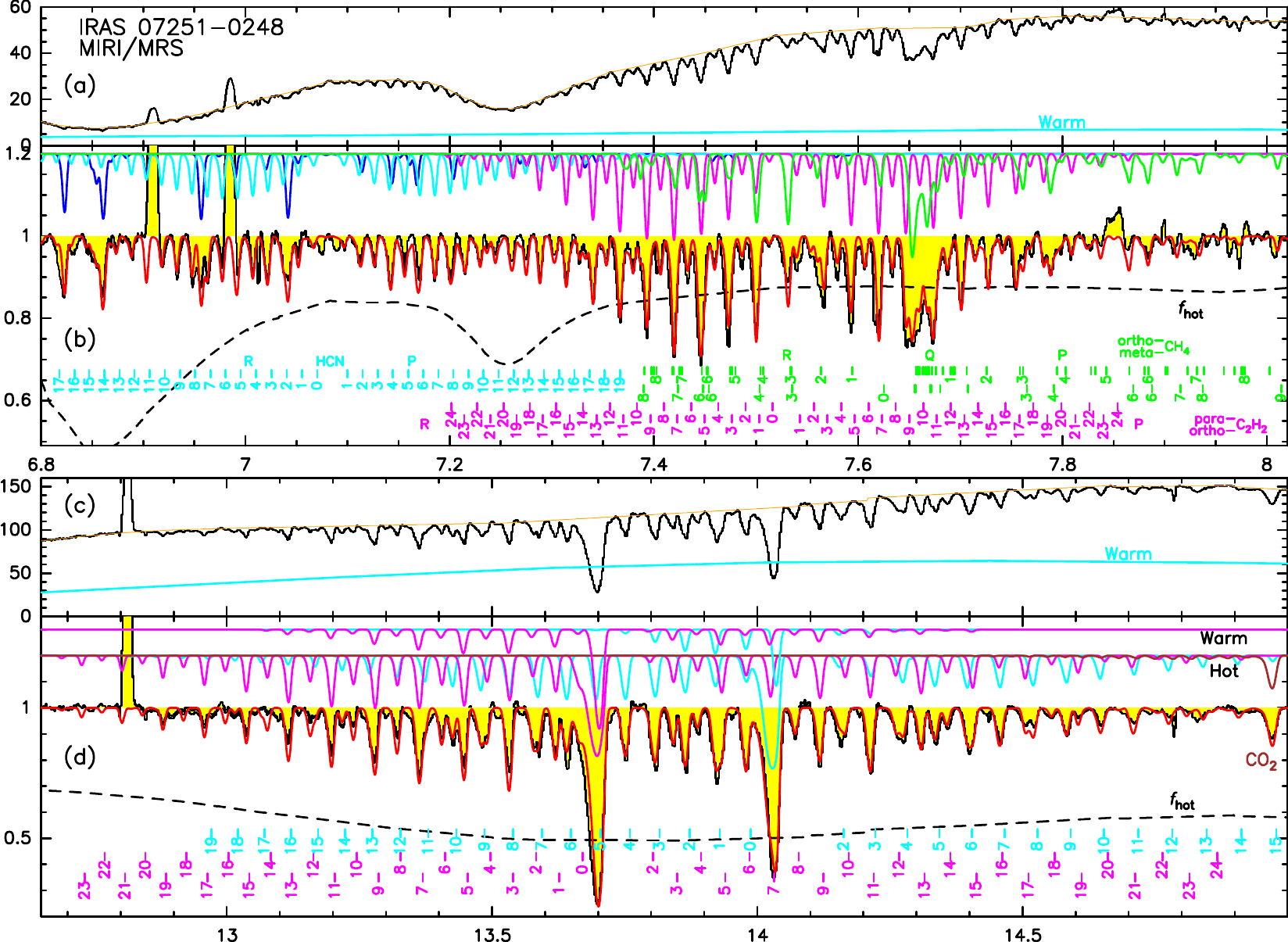}
\caption[]{HCN, C$_2$H$_2$, and CH$_4$ bands in IRAS~07251$-$0248 observed
  with {\it JWST} MIRI/MRS, and model fit. Abscissae are rest wavelengths
  in $\mu$m. Panels a and c show the observed
  spectra around 7.5 and 14\,$\mu$m (flux densities in mJy),
  with the orange line indicating
  the adopted continuum level (i.e., the baseline)
  and the light-blue curve showing the
  contribution to the continuum by the $W_C$. Panels b and d
  show the continuum normalized spectrum, with the
  red line showing the model result  
  and the dashed line indicating the covering factor of the
  hot component ($f_{\mathrm{hot}}$).
  The contributions to the model by HCN, C$_2$H$_2$, CH$_4$,
  and CO$_2$ (and their transition labels)
  are indicated in light-blue, magenta, green,
  and brown, respectively
  (and vertically shifted for clarity).
  Note that all absorption features are
  blueshifted relative to the labels.
}
\label{fighcnc2h2}
\end{figure*} 

\begin{figure*}[h]
   \centering
\includegraphics[width=18.3cm]{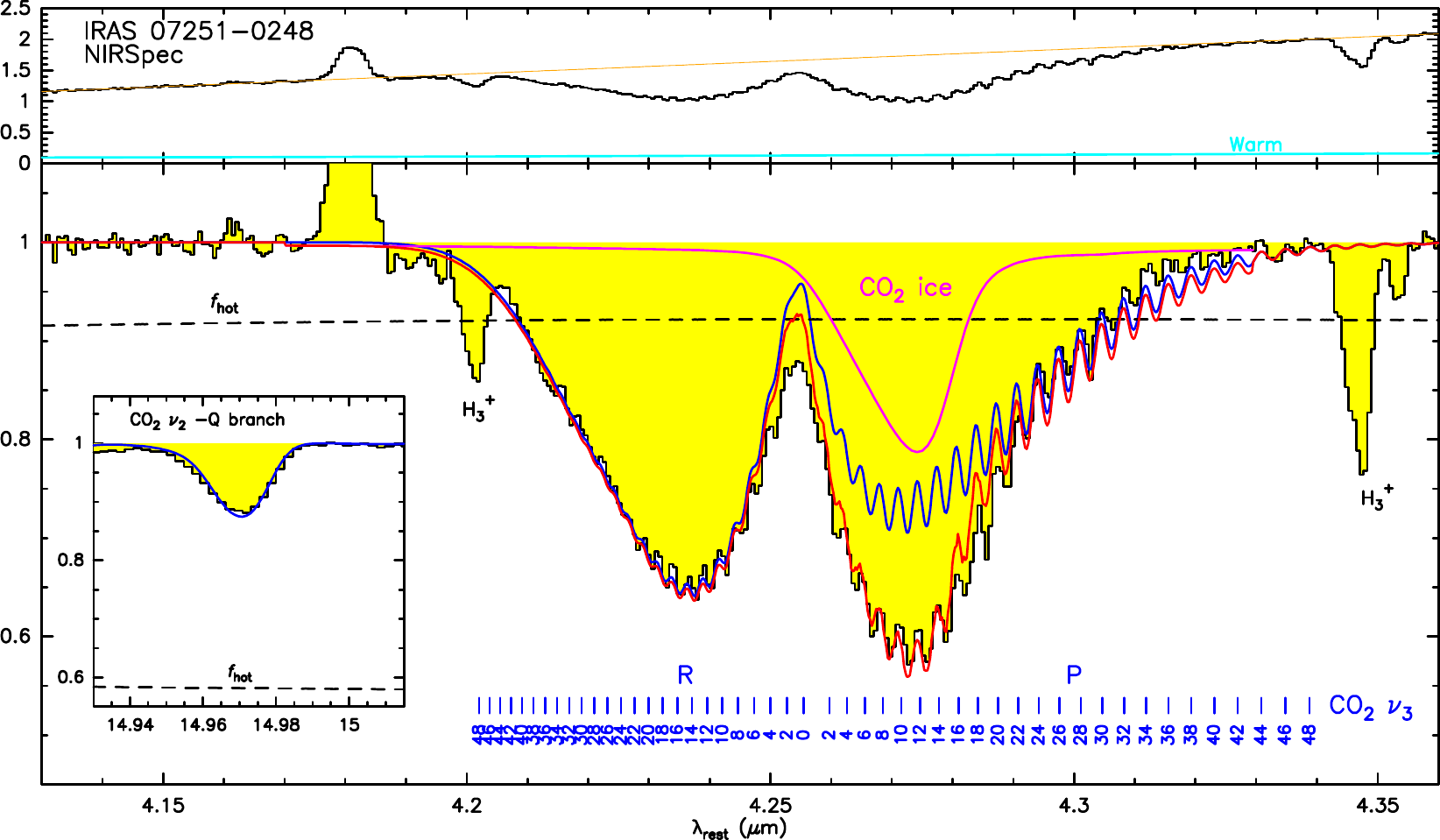}
\caption[]{CO$_2$ $\nu_3$ and $\nu_2$ bands in IRAS~07251$-$0248 observed
  with {\it JWST} NIRSpec, and model fit. The upper panel a shows the observed
  spectrum around 4.25\,$\mu$m (flux densities in mJy),
  with the orange line indicating
  the adopted continuum level (i.e., the baseline)
  and the light-blue curve showing the
  contribution to the continuum by the $W_C$. The lower panel
  shows the continuum normalized spectrum, with the
  blue line showing the model result for the $H_C$
  and the dashed line indicating the covering factor of the
  hot component ($f_{\mathrm{hot}}$).
  We have also included the required contribution to the absorption
  by CO$_2$ ice ($\exp\{-\tau_{\mathrm{ice}}\}$ in magenta); red is the
  resulting spectrum.
  The insert shows the Q branch of the CO$_2$ $\nu_2$ band at
  $\approx15$\,$\mu$m, with the blue line also indicating the model
  result for the $H_C$.
}
\label{figco2}
\end{figure*} 

The {\it JWST} MIRI/MRS spectrum of IRAS~07251$-$0248 at $6.8-8.0$\,$\mu$m in
Fig.~\ref{fighcnc2h2}a shows a forest of absorption lines
that are identified with the gas-phase HCN 2$\nu_2$ 7.1\,$\mu$m,
C$_2$H$_2$ $\nu_4+\nu_5$ 7.5\,$\mu$m, and CH$_4$ $\nu_4$ 7.7\,$\mu$m bands
\citep{gber25}.
Spectroscopic parameters for these species were
  obtained from the HITRAN2020 \citep{gor22} and ExoMol \citep{chu20,yur24}
  databases. The strengths of the individual lines (panel b) are
comparable to those of the H$_2$O $\nu_2$ lines, which is surprising given
that these bands are intrinsically weak. For the HCN 2$\nu_2$ 
and C$_2$H$_2$ $\nu_4+\nu_5$ bands, the
Einstein coefficients for absorption of radiation ($B_{lu}$) are
$\sim10^6$ and $\sim2.5\times10^6$\,cm$^2$\,erg$^{-1}$\,s$^{-1}$,
respectively, while the $B_{lu}$ values for the backbone-backbone
transitions of the H$_2$O $\nu_2$
lines are factors of $\sim5$ and $\sim2$ higher.

We have used our model for the $H_C$, which dominates the continuum at
these wavelengths, to fit these bands, with the goals of checking the
excitation predicted by the continuum model
and deriving the abundances of these species. Our model uses, as for H$_2$O,
a $\lambda$-dependent covering factor $f_{\mathrm{hot}}$
as defined in eq.~(\ref{fhot}) (Section~\ref{sech2oband}).
  The dilution of the molecular absorption at $7-8$\,$\mu$m
  is also due to the ``hidden'' 230\,K component (Fig.~\ref{sed}).
The best-fit model is shown in Fig.~\ref{fighcnc2h2}b.
To explain the band strengths
we require extraordinary abundances relative to H$_2$O,
indicating an extreme C-rich chemistry:
$\mathrm{[HCN]/[H_2O]}\approx0.56$, $\mathrm{[C_2H_2]/[H_2O]}\approx0.54$, 
and $\mathrm{[CH_4]/[H_2O]}\approx0.22$. These are within a factor 2 consistent
with the values inferred by \cite{gber25}.

The HCN $\nu_2$ 14\,$\mu$m and C$_2$H$_2$ $\nu_5$ 13.7\,$\mu$m bands are also
clearly detected (Fig.~\ref{fighcnc2h2}c). These bands have $B_{lu}$ values
higher than the HCN and C$_2$H$_2$ $7.1-7.5$\,$\mu$m bands by factors of
$5-10$, and thus the P and R branch lines from the $H_C$ at
  14\,$\mu$m, predicted by our model for the $\sim7.5$\,$\mu$m bands,
  would be expected to be strong.
However, the $W_C$ continuum emission at 14\,$\mu$m,
here dominated by the $W_C$ 123\,K component (Table~\ref{tab}),
is expected to account
for about half of the observed continuum (Figs.~\ref{sed} and \ref{fighcnc2h2}c)
and thus the 14\,$\mu$m absorption from the $H_C$ is diluted with
$f_{\mathrm{hot}}\sim0.5$ (Fig.~\ref{fighcnc2h2}d).
This dilution has the effect of reducing the predicted
  14\,$\mu$m high-$J$ molecular absorption troughs at levels comparable with the
  observations.
  Using a common model for the $H_C$, our composite continuum model thus
  yields consistent results for the 7.5 and 14\,$\mu$m bands,
  which is required as the high-J lines of the R and P branches in both bands
  are produced by the same gas.

However, the Q branches of both 14\,$\mu$m bands are saturated, and
  the model for the $H_C$ alone cannot reproduce them as they absorb more than 50\% of
the continuum. We have therefore added a model for the $W_C$
in Fig.~\ref{fighcnc2h2}d that accounts for the remaining Q branch absorption,
using a covering factor for the $W_C$ of $f_{\mathrm{warm}}=1-f_{\mathrm{hot}}$
(note that the $C_C$ contributes negligibly to the 14\,$\mu$m continuum,
Fig.~\ref{sed}). The absorbing column densities of both HCN and C$_2$H$_2$
in the $W_C$ are $10^{17}$\,cm$^{-2}$, generating absorption in the Q branches
and in the low-$J$ lines of the P and R branches.
The low-$J$ lines of the HCN 14\,$\mu$m
band have additional contribution by the low-excitation shell in front of the $H_C$,
and indeed absorption is seen up to $\sim-500$\,km\,s$^{-1}$ in the uncontaminated
HCN $\nu_2$ R4 line (this has the effect of lowering the apparent $T_d$). 
The fit to the 14\,$\mu$m bands tends to overestimate the absorption in 
high-excitation lines, most likely reflecting the uncertainties of
$f_{\mathrm{hot}}$.

From the results for the HCN abundance found here and by \cite{gber25},
and those for the HCO$^+$ abundance by \cite{spe25}, a
very high [HCN]/[HCO$^+$] abundance ratio of $\sim100$ is obtained in IRAS
07251$-$0248 E. This result can be related to
the variety of HCN/HCO$^+$ rotational line ratios observed 
galaxies, and widely discussed in the literature
\citep[e.g.,][]{kri08,ima09,izu16}. 
While a high HCN/HCO$^+$ line ratio was first
attributed to the chemistry involved in X-ray dominated regions
around AGNs, further observations showing high ratios in non-AGN
galaxies and low ratios in AGN
have dismissed this relationship \citep[e.g.,][]{cos11,pri20}.
Recently, \cite{nis24} have identified the highest HCN/HCO$^+$ line
ratios in shell-like structures kinematically associated
with outflows or inflows, and their chemical models indeed indicate
a strong enhancement of [HCN]/[HCO$^+$] in shock conditions.
The nuclear conditions in IRAS~07251$-$0248\,E are more extreme
than those modeled by \cite{nis24}, which
are appropriate for the more extended regions probed by the rotational lines,
but their shock models illustrate the high [HCN]/[HCO$^+$] ratio that can be
attained in the outflow associated with the $H_C$. Nevertheless,
the additional effect of cosmic rays in enhancing
the ionization rate and processing the carbonaceous grains and PAHs
are most likely required to explain
the extreme $\mathrm{[HCN]/[H_2]}\sim4\times10^{-5}$ we find in the
nucleus of IRAS~07251$-$0248\,E.

Finally, we have also applied the $H_C$ model to the
CO$_2$\,$\nu_3$ ($4.25$\,$\mu$m) and $\nu_2$ ($14.97$\,$\mu$m) bands
(Fig.~\ref{figco2}). The best fit model to both the CO$_2$\,$\nu_3$
R branch and the $\nu_2$ Q branch requires
$\mathrm{[CO_2]/[H_2O]}\approx0.03$ at the $H_C$ photosphere, 
with an additional contribution by the surrounding outflowing
shell with $\mathrm{[CO_2]/[H_2O]}\approx0.014$. The fit, however,
grossly underestimates the strength of the CO$_2$\,$\nu_3$
P branch up to $J\sim20$, so we added foreground CO$_2$ ice with
$\mathrm{H_2O:CO_2=10:1}$ \citep{ehr97,roc22}.
The resulting modeled spectrum appears to
overestimate the absorption
in the high-$J$ lines of the $\nu_3$ P branch,
but fits both the $\nu_3$ R branch
  at 4.23\,$\mu$m and the $\nu_2$ Q branch at 15\,$\mu$m rather well.
  This match validates our fit to the $H_C$ continuum.

\end{appendix}

\end{document}